\documentclass{article}
\usepackage{apacite}
\usepackage{natbib}
\usepackage{graphicx} 
\usepackage[a4paper, total={6in, 9in}]{geometry}
\usepackage{hanging}
\usepackage{enumitem}
 \usepackage{array}
\usepackage{setspace}
\usepackage{multirow}
\usepackage{bbding}
\usepackage{color}
\definecolor{lightgray}{gray}{0.85}

\usepackage{float}
\usepackage{xcolor}
\usepackage[normalem]{ulem}
\usepackage{tikz}
\usetikzlibrary{shapes.geometric, arrows.meta, positioning, fit, backgrounds}

\newcommand\greybox[1]{%
  \vskip\baselineskip%
  \par\noindent\colorbox{lightgray}{%
    \begin{minipage}{\textwidth}#1\end{minipage}%
  }%
  \vskip\baselineskip%
}

\title{Generative AI use in Statistical Research: \\ A Literature Review and Code Generation Case Study}

\author{Natalie Morosin, Adel Ahmadi Nadi,   and    Michael P. Wallace  \vspace{0.2cm}\\
{\small {Department of Statistics and Actuarial Science, University of Waterloo, Waterloo, Canada.}}}
\date{}

\begin{document}

\maketitle

\begin{abstract}

Generative artificial intelligence (GenAI) is a large language model (LLM) that has the ability to generate media based on user-provided prompts. Given the demonstrated capabilities of models such as ChatGPT in information synthesis and programming, there is growing interest in their potential role within the research process. However, little work has evaluated recent GenAI models for research tasks in the domain of statistical research. This case study examines GenAI as a tool for developing a literature review and translating methodology from academic papers into code, for the topic of dynamic treatment regime (DTR) estimation via the dynamic weighted ordinary least squares (dWOLS) approach. Specifically, we utilize ChatGPT-5 and ScholarAI (Sept-Nov 2025 release) in the processes of identifying relevant sources for the literature review, creating summaries of papers, identifying gaps in research, and R code generation to implement methodology. Our findings show that current GenAI models lack the depth and contextual understanding required to accomplish these tasks without careful prompting and supervision of a knowledgeable researcher. Nonetheless, GenAI has potential to increase efficiency of tasks which take advantage of its search and summarization abilities, as well as basic code debugging and algorithm formation. We demonstrate that under a knowledgeable guide, GenAI can function as a research tool, but not as a substitute for methodological expertise.

\end{abstract}

\noindent \textbf{\textit{Keywords: }} Generative artificial intelligence;  large language model; AI assistant; ChatGPT; ScholarAI

\section{Introduction}

    Generative artificial intelligence (GenAI) is a type of large language model (LLM) which can generate text and other media in response to user-provided prompts. GenAI has become extremely popular in the last few years, and one of the most well-known models is ChatGPT trained by OpenAI. ChatGPT is trained on publicly available information from the internet, information from third parties partnered with OpenAI, and information from human trainers and researchers \citep{OpenAI2025}. Other popular models include Claude (Anthropic), LLaMa (Meta) and Gemini (Google). 

    While GenAI has begun to be adopted into many different industries, it has been questioned whether GenAI can be used in research settings. In a study that gathered medical researchers’ perspective on the use of AI chatbots in the scientific process, it was found that 69\% of researchers were interested in training to learn how to use AI, and 60\% of respondents felt that AI chatbots would be helpful in conducting literature reviews \citep{Ng2025}. Others point out that LLMs and GenAI can reduce time in screening publications for literature reviews, and provide support when synthesizing research or identifying research gaps \citep{DelgadoChaves2025, Galli2025, Schryen2025}.
    
    In addition to literature reviews, another potential use of GenAI in the research space is in generating programming code. While it is increasingly standard practice to publish source code that employs a method alongside its published paper, there are many cases where code is not available. Thus, another practical use of GenAI is to translate methods from scientific research papers, so researchers can use the method in their own work. This also supports the goal of increasing accessibility and reproducibility of methods in research.

    Despite the potential utility, there is very little work evaluating the use of GenAI in the context of statistical methodological research. In this paper, we will evaluate the use of GenAI to conduct a statistical literature review and generate code in the context of dynamic treatment regime (DTR) estimation methods for precision medicine. For this evaluation, we use an existing literature review on the same topic and code developed by domain experts as benchmarks for assessing the performance of GenAI tools.  We note that, as this work is necessarily limited to GenAI tools available at the time the research was conducted, newer versions are already available. Nevertheless, we believe this work offers valuable insights into the use of such tools in the past, present, and future. These insights may be particularly valuable for early-career researchers, who are more likely to adopt these tools but may have more limited domain-specific expertise.

\subsection{GenAI Use in Scientific Literature Reviews}

Literature reviews are fundamental to the research process, but can be resource intensive. With their proven capabilities in finding and synthesizing information, GenAI may seem like a natural tool in creating such reviews. Several researchers have recognized this potential. Scherbakov et al. (2024) found that the most popular model in studies on evaluating GenAI in literature reviews is ChatGPT (OpenAI), using the then-recent versions GPT-3.5 and GPT-4. The next popular models are LLaMa/Alpaca (Meta), Gemini (Google), Claude (Anthropic).
        
        Some works describe using chatbots like ChatGPT, including prompts, example responses, and recommendations for researchers who wish to employ the model in their own work \citep{Schryen2025, Susarla2023}. These works show promising results of GenAI for expanding scope of search and improving efficiency of researchers, however, the authors discuss limitations of using these models, including hallucinations, lack of depth and consistency when producing summary results, and ethical concerns.

        A “hallucination” occurs when content is generated that seems reasonable but contains factually incorrect information \citep{Gu2024}. There are many factors that contribute to GenAI hallucinating, for example, GenAI is trained on both factual and fictional data and is designed to produce the most statistically probable response, rather than the most factual response. Many researchers observed hallucinations while utilizing GPT-3.5: \citet{Winberg2025} focused on literature retrieval and found that 42\% of citations were not real and \citet{Haman2023} found that 66\% of retrieved articles were completely fabricated. \citet{Susarla2023} also highlighted hallucinations and entirely fictitious citations when evaluating the use of ChatGPT for summarizing texts and creating literature reviews. They also point out that GenAI has an inherent lack of context and scholarly depth, as well as missing citations which puts the credibility of the information up to question. GPT-4 has also been explored for usage in research. \citet{Mostafapour2024} compared two literatures reviews, one created by a human and one created using ChatGPT (GPT-4). They found that ChatGPT showed an impressive breadth of knowledge and extremely fast response time but limited contextual and deep understanding. They highlight how the human literature review has the advantage of transparency (i.e. providing literature search methods and rationalization behind decisions), whereas the nature of ChatGPT results in uncertainty as to how the model processes prompts, collects information, and generates responses. 

        Additionally, some studies describe how GenAI and LLMs can be used to automate a particular phase of the review. For instance, when evaluating the use of AI searching and screening relevant papers, one common approach is to use ChatGPT API through custom programming scripts \citep{Guo2024, Galli2025, Issaiy2024}.
        
        While OpenAI’s ChatGPT has received much attention, there are many other tools that have been evaluated in published literature. Specifically, GenAI tools such as LitMaps, Scholarcy, Scite, ResearchRabbit, ScholarAI, Elicit, and Consensus were developed specifically to support research-related tasks. \citet{Kumar2025} had several doctoral students use research-specific GenAI tools to create a literature review, in which they found that GenAI can support efficiency, discovery steps, making connections, and recognizing gaps in literature. However, much of the output from GenAI lacked accuracy and contained hallucinations. Regardless, other articles focus on recommendations for how GenAI and AI-based tools can supplement each step of the literature review and prompting techniques, including examples of tools and comparisons to common non-AI based tools \citep{Bolanos2024, Robledo2023, Wagner2022}.
        
        Current research that evaluates using GenAI for literature reviews often targets the subject of information systems \citep{Ngwenyama2024, Schryen2025, Wagner2022}. Another common approach is to use GenAI to perform literature reviews on the topic of using GenAI to perform literature reviews \citep{Adel2025, DelgadoChaves2025, Mozelius2024, Scherbakov2025}. \citet{Gwon2024} discusses using ChatGPT for literature searches for healthcare research, and uses a grading system to evalaute the accuracy of responses.

        \subsection{GenAI Use in Code Generation}
        
        Researchers have investigated the use of LLMs and GenAI agents in generating code, but like summarizing and generating text, GenAI’s code is subject to hallucinations and other errors. In one study by \citet{Vaithilingam2022}, participants completed python programming tasks with the aid of Microsoft Copilot. It was found that the generated code introduced new bugs that participants found difficult to fix, resulting in no improvement in task completion or success rate. However, it is noted that GenAI provided a useful starting point or an alternative to searching online. Another study showed that GenAI can significantly boost productivity in coding, especially for beginners \citep{Yang2024}. Continuing, there are papers which develop benchmarks to measure LLM’s ability to produce code, for example, in data science tasks and Python programming tasks \citep{Jing2025, Zhuo2025}.
        
        In addition to exploring GenAI’s capability to complete programming tasks, researchers have explored using GenAI in generating code based on methodology described in an academic paper. \citet{Bibal2025} found that AI can significantly reduce time spent reproducing scientific research. However, AI has limited scientific understanding and struggles in tasks that require logic and knowledge of underlying assumptions which might not be explicitly stated in the paper. 
        
        OpenAI’s PaperBench was a study conducted to test if various AI agents could replicate methodology and experiments of machine learning papers, specifically in generating code \citep{Starace2025}. The agents tested were GPT-4o, o1, o3-mini, DeepSeek-R1, Claude 3.5 Sonnet, and Gemini 2.0 Flash, mainly in writing Python code. Results found that models outperformed humans in initial stages, but humans outperformed in the long-term. Similar to the previous paper, they concluded that LLMs can grasp some methodology and produce code, but lack the accuracy and completeness of human baselines \citep{Starace2025}. In a similar setting, researchers found that using a staged reasoning approach improved success, and required minimal edits to recieve executable code \citep{Seo2025}.
        
        Other research topics have been explored, for instance, \citet{Bersenev2024} evaluated using ChatGPT (GPT-4) to produce python code from the methodology sections of a high-impact publication in a biological topic. Another study tasked ChatGPT (GPT-4) to code algorithms in MATLAB based on a paper in geotechnical engineering \citep{Sharma2025}. Both groups found ChatGPT had difficulties producing functional code and required a human with domain-knowledge to debug, verify code methods, and judge results. 
        
        The book “Leveraging GenAI for Machine Learning Education in Public Health: ChatGPT and R” discusses using GenAI to program in R, focusing on machine learning and public health contexts. In particular, the author discusses how ChatGPT can be used to support users and integrate into workflow \citep{Leung2025}.

    \subsection{Goals}
        
        The focus of this case study is to take the perspective of a statistical methodology researcher and evaluate how GenAI can be used as a tool in the creation of a literature review and for code generation. We aim to give insight into the following questions:
        
        \begin{enumerate}
            \itemsep0em 
            \item How can generative artificial intelligence be used to assist in a statistical literature review?
            \item How can generative artificial intelligence be used to develop code based on a statistical research paper?
        \end{enumerate}

We will evaluate strengths and limitations of GenAI for these two tasks, and make recommendations for researchers who might want to incorporate GenAI into their research workflow.

        The remainder of the paper is organized as follows: Section 2 evaluates the use of GenAI inc creating a literature review, including conducting a literature search, determining relevancy of the identified papers, generating summaries of the papers, and identifying gaps in research. Section 3 explores using GenAI to translate methodology from a research paper into code. Section 4 concludes the paper with a discussion of results, limitations, and future research. Finally, the Appendix includes further details for Section 2 and 3, including prompts, GenAI responses, and sample code. 

\section{Generative AI Use in a Statistical Methodology Literature Review}
        
    \subsection{Methodology}
    
        We will focus on replicating a literature review that has already been conducted. The literature review ``Recent advances in doubly-robust weighted ordinary least squares techniques for dynamic treatment regime estimation'' \citep{Nadi2025}, discusses developments in doubly-robust DTR estimation through a method called dynamic weighted ordinary least squares (dWOLS, \citeauthor{Wallace2015}, \citeyear{Wallace2015}).
        
        DTRs are a technique employed to operationalize precision (or personalized) medicine, in which treatment decisions are individualized based on patient-level characteristics. Identifying the \emph{optimal} DTR (that is, the sequence of treatment decisions that improve some clinical outcome) is an important challenge in precision medicine, with numerous methods developed for this purpose. dWOLS is one such method, characterized by ease of implementation through regression-based methods, and robustness to model misspecification.
        
        Initially developed for a binary treatment decision (e.g., drug A or B) and a continuous outcome (such as a continuous measure of health, or symptom severity), multiple advancements of the dWOLS method have been developed. This includes extensions to more outcome types (e.g., categorial, continuous, and time-to-event data) and adaptations when assumptions are violated (e.g., presence of measurement error in observations, presence of interference between individual observations). The Nadi review provides a comprehensive summary of dWOLS-based DTR estimation methods, including theoretical extensions and practical implementations on real-world problems.

At the time our GenAI case study was conducted, this literature review had not yet been formally published. However, a version was available on arXiv, and therefore is technically accessible to GenAI models. Nonetheless, because this literature review was completed by researchers without the use of generative AI, it provides a useful reference point to compare with GenAI's output. We will refer to this literature review as the ``Nadi review".

            To analyze GenAI's ability to replicate the Nadi review, we will evaluate if GenAI can be used as a tool in completing the following tasks: 1. Identify relevant sources (extensions of dWOLS methodology and papers where dWOLS was applied in real data analysis); 2. Synthesize sources (summarize and identify relevant papers); 3. Identify gaps in research; and 4. “Write” a literature review. The assessment of the relevance of papers, and the accuracy of summaries and research gap identification, was conducted by this review's co-authors, both with extensive expertise in the subject matter. 

        The first step in conducting our AI-assisted review is deciding which GenAI tool we will use. As discussed in Section 1, there are two domains in which GenAI models reside: “general” chatbots, and research-specific models.
        
        For chatbots, we considered widely used models including ChatGPT, Claude, and Gemini. Each of these models offer capabilities which are relevant to our goal of producing a literature review, with similar strengths and limitations. We selected ChatGPT due to its prominence and relevancy in current literature. GPT-5 was released in August 2025 (this project began in September 2025) and was reported to greatly outperform previous models in terms of quality of responses, including a significant reduction in hallucinations \citep{OpenAI2025}. The introduction of OpenAI’s reasoning model \emph{GPT Thinking} promises improvements in the reasoning capabilities of the models, as it is trained to try different strategies and recognize its mistakes before responding to the user \citep{OpenAI2025}. 
        Additionally, ChatGPT \emph{Deep Research}, launched in February 2025, is a model that can perform complex tasks by conducting multi-step research on the internet. The full Deep Research model is only available on premium plans. Continuing, OpenAI offers ChatGPT “GPTs” which are additional tools built by third parties which use OpenAI API and the ChatGPT interface. Combined, this presents an avenue of ChatGPT which has been unexplored in current published research and may present improved results when aiding the creation of a literature review. ChatGPT will be utilized with the subscription-based “Plus” plan, to reduce limitations on using GPT-5, unlock the Deep Research model, and for increased Portable Document Format (PDF) uploading capabilities.
        
        In addition to general-purpose chatbots, a range of research-specific GenAI tools have been developed to support researchers. Some take the form of academic search tools without a conversational tool, while others utilize the familiar chatbot interface. These models focus on retrieving information from academic databases rather than relying primarily on pre-trained data or a broad web search. They also contain features for uploading papers for analysis or summarization. For the purposes of this project, and to allow direct comparison to ChatGPT, we will focus on a research-specific GenAI in the form of a chatbot. Here, we considered ScholarAI, Elicit, Consensus, and Semantic Scholar. ScholarAI stood out as it seemed to have the widest breadth of search, as well as advertising a capability to deeply analyze academic papers and create literature reviews. The model was specifically created to combat the high hallucination risk of LLMs such as ChatGPT when searching for papers by sourcing from strictly academic databases. ScholarAI uses a chat-based interface to search papers from sources including PubMed, arXiv, and Google Scholar through web-indexing. ScholarAI also exists as a ChatGPT ``GPT". Thus, ScholarAI can be accessed on its own platform (which will be subsequently referred to as “ScholarAI”) and the GPT version of ScholarAI can be accessed through ChatGPT (“ScholarAI GPT”) which utilizes the most recent version of GPT. In our usage, ScholarAI GPT will utilize GPT-5 while ScholarAI was built on GPT-4 API. Therefore, ScholarAI with the subscription-based “Premium” plan was chosen, to access unlimited PDF uploads and to provide a sufficient number of `credits' to use for prompts.
        
        We used premium versions of these models primarily to avoid limitations in number of responses and document uploads. Both these models have versions which can be accessed at no cost, but the paid plans provide extended access to the more recent and powerful models.
        
        To summarize, we will employ the following GenAI models: ChatGPT (GPT-5), ScholarAI GPT (GPT-5), and ScholarAI (September – November 2025 release). For the remainder of this section, the term ``GenAI" will be used to collectively denote these three models in the listed release versions, as applied within the specific use case examined in this study. For both ChatGPT and ScholarAI, we turned off the options to share our conversation data. GPT-5's Thinking model can be toggled on, off, or ``auto" by the user. ChatGPT was set to ``auto-thinking" when searching for sources, and set to ``thinking" for summarizing, organization, and identifying gaps.

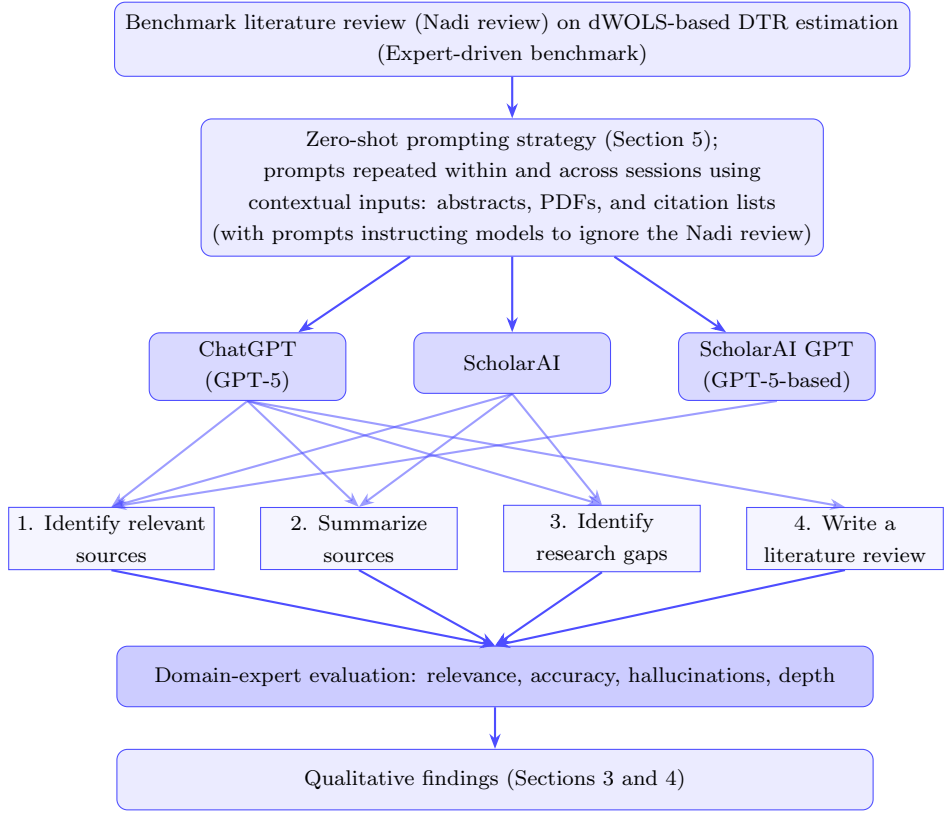
\begin{figure}[H]
    \centering
    \begin{tikzpicture}[
        node distance=0.55cm and 0.6cm,
        every node/.style={font=\footnotesize, align=center},
        block/.style={rectangle, rounded corners, draw=blue!70, fill=blue!8,
                      minimum width=3.0cm, minimum height=0.8cm, inner sep=4pt},
        model/.style={rectangle, rounded corners, draw=blue!70, fill=blue!15,
                      minimum width=2.6cm, minimum height=0.8cm, inner sep=3pt},
        task/.style={rectangle, draw=blue!70, fill=blue!4,
                     minimum width=2.6cm, minimum height=0.7cm, inner sep=3pt},
        eval/.style={rectangle, rounded corners, draw=blue!70, fill=blue!20,
                     minimum width=10.0cm, minimum height=0.8cm, inner sep=4pt},
        arr/.style={-{Stealth[length=2mm]}, draw=blue!70, thick}
    ]
        \node[block] (ref) {Benchmark literature review (Nadi review) on dWOLS-based DTR estimation\\(Expert-driven benchmark)};
        \node[block, below=of ref] (prompt) {Zero-shot prompting strategy (Section~\ref{appendix});\\
prompts repeated within and across sessions using\\
contextual inputs: abstracts, PDFs, and citation lists\\(with prompts instructing models to ignore the Nadi review)};
        \node[model, below=1.0cm of prompt, xshift=-3.5cm] (gpt) {ChatGPT\\(GPT-5)};
        \node[model, below=1.0cm of prompt] (sa)  {ScholarAI};
        \node[model, below=1.0cm of prompt, xshift=3.5cm]  (sgpt){ScholarAI GPT\\(GPT-5-based)};

        \node[task, below=1.4cm of gpt, xshift=-1.8cm] (t1) {1. Identify relevant\\sources};
        \node[task, right=of t1] (t2) {2. Summarize\\sources};
        \node[task, right=of t2] (t3) {3. Identify\\research gaps};
        \node[task, right=of t3] (t4) {4. Write a\\literature review};

        \node[eval, below=1.0cm of t2, xshift=1.8cm] (expert)
            {Domain-expert evaluation: relevance, accuracy, hallucinations, depth};
        \node[block, below=of expert, minimum width=10.0cm]
            (out) {Qualitative findings (Sections 3 and 4)};

        \draw[arr] (ref) -- (prompt);
        \draw[arr] (prompt) -- (gpt);
        \draw[arr] (prompt) -- (sa);
        \draw[arr] (prompt) -- (sgpt);

        \foreach \m in {gpt,sa}{
            \foreach \t in {t1,t2,t3}{
                \draw[arr, opacity=0.55] (\m.south) -- (\t.north);
            }
        }

         \draw[arr, opacity=0.55] (sgpt.south) -- (t1.north);
         \draw[arr, opacity=0.55] (gpt.south) -- (t4.north);
        
        \draw[arr] (t1.south) -- (expert.north);
        \draw[arr] (t2.south) -- (expert.north);
        \draw[arr] (t3.south) -- (expert.north);
        \draw[arr] (t4.south) -- (expert.north);
        \draw[arr] (expert) -- (out);
    \end{tikzpicture}
    \caption{Overview of the case-study workflow.}
    \label{fig:workflow}
\end{figure}

            Responses from GenAI are subject to extreme variability depending on the prompts used and context already provided to the model. Prompt engineering describes specially crafting inputs to obtain relevant and accurate results. In this case study, we employed \emph{zero-shot} prompting, in which the given prompt does not contain any examples or demonstrations. This was intended to replicate the most straightforward manner in which a researcher may employ GenAI in their research workflow. As well, identical prompts were used across all three models to allow for direct comparison. Prompts were used multiple times in the same session, and across new sessions to derive a sense of how responses varied depending on the context of the conversation. Figure~\ref{fig:workflow} provides an overview of the case-study workflow, from prompt design and GenAI interaction through domain-expert evaluation.

            It is important to note that we cannot perfectly prevent ChatGPT or ScholarAI from accessing the arXiv submission of the Nadi review. To combat this, we used prompts with explicit instructions to ignore this paper, and the visible “chain-of-thought” from ChatGPT’s Thinking model was monitored, as this gives information as to where the model sourced its response from. Nonetheless, the nature of these models may result in some hidden biases that we could not prevent.

            For certain tasks, we uploaded PDF files of published papers to analyze the model's capabilities with access to the full-text. We verified that our work was aligned with publisher rules and guidelines about uploading of published works. We also ensured our work was consistent with responsible GenAI use, which includes literature synthesis (e.g., summarization) and programming support. Accordingly, any paper-derived information reported in this study is drawn exclusively from publicly available sources, such as abstracts or open-access articles.
    
\subsection{Results}
    \subsubsection{Identifying Sources}

        We first use GenAI to find as many relevant sources as possible on the topic of dWOLS methodology for DTR estimation. Specifically, we want to find the papers cited in the Nadi review and any additional sources incorrectly omitted from Nadi's review which should be included in a literature review of this topic.
    
        Typically, identifying relevant literature would be done by searching academic databases. Thus, we will use the various GenAI models as our “search tool” using zero-shot prompting. We will request papers that use dWOLs methodology and extend it, and papers that apply dWOLS method for real data analysis. We will ask this prompt multiple times on each model, both in concurrent and new sessions. Then, we will make observations on the sources found, including any errors or hallucinations. 

        The following is an example of a prompt and follow-up prompt. The entire list of prompts used can be found in the Appendix section \ref{sources}.
        \greybox{  \textbf{Prompt 1. a)} Find articles that incorporate and extend the methodology from ``Doubly-robust dynamic treatment regimen estimation via weighted least squares" by Wallace \& Moodie.

        \textbf{b)}  Find more articles that incorporate and extend the dWOLS methodology from ``Doubly-robust dynamic treatment regimen estimation via weighted least squares" by Wallace \& Moodie and provide title, date published, author names, and abstract.}
        
\begin{figure}[H]
            \centering
            \includegraphics[width=0.85\linewidth]{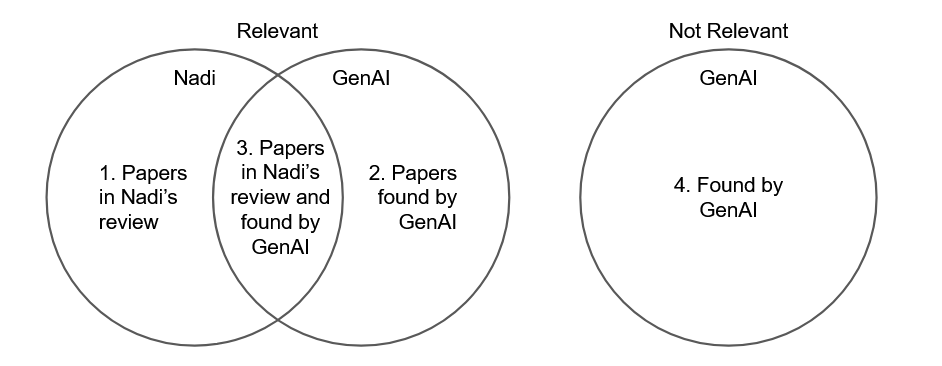}
            \caption{A Venn diagram visualizing the papers we will identify in this case study. The papers will either be relevant or not relevant, and with four subcategories: 1. Relevant papers which are in Nadi's review, 2. Relevant papers which are found by GenAI, 3. Relevant papers which are in Nadi's review and found by GenAI, 4. Not relevant papers found by GenAI.}
            \label{fig:venn}
        \end{figure}

        We can classify the types of papers gathered into two broad categories: ``relevant", and ``not relevant". Papers which are relevant extended dWOLS methodology, and/or applied dWOLS in real data analysis. Papers that are not relevant do not discuss dWOLS to any substantive level and should therefore not be mentioned in a literature review of the topic. Figure \ref{fig:venn} visualizes the set of papers identified across the Nadi review and/or by GenAI in categories.

         We are particularly interested in papers which appear in category 2: Relevant papers which are found by GenAI but do not appear in the Nadi review. Here, we can identify if GenAI was a beneficial tool through identifying additional relevant sources. Conversely, category 4 captures papers that GenAI recommended for inclusion but were ultimately deemed irrelevant. This category will demonstrate instances where the model misunderstood a paper's connection to the topic.

        All GenAI succeeded in producing relevant papers on the target topic, although the models varied in how many papers they were able to identify (with further prompting eliciting either no additional papers, or duplicates of those already found). In addition to open-access articles, all models identified papers behind paywalls. In the latter case, GenAI sourced these papers having only access to the title, abstract, and metadata. 
            
        Over our three models (ChatGPT, Scholar AI, Scholar AI GPT) we identified 40 unique papers. See Appendix section \ref{papers} for details on these papers and their classification. In a manual review of these papers, we determined 26 to be relevant, and the remaining 14 not relevant. Looking at each model individually, ChatGPT gathered the largest number of unique sources (26) and the largest proportion of relevant articles, 22 out of 26. ScholarAI identified 21 unique sources where 12 were relevant; ScholarAI GPT found 20 unique sources where 14 were relevant. ChatGPT identified the largest number of articles which appeared in the Nadi review, along with 8 new relevant articles (not contained in Nadi review). ScholarAI identified 2 papers that were about an entirely unrelated subject, and was the only model to do so. Details on each model's performance can be found in the Appendix section \ref{sources}. 

        Only one paper which appeared in the Nadi review was not identified by any model. We believe this paper was not found because it did not explicitly use any of the keywords which GenAI might be looking for (e.g. “dWOLS”). A comparison of model performance in the source-identification task is provided in Table~\ref{tab:source_summary}. 

             \begin{table}[H]
        \setlength{\tabcolsep}{3pt} 
\renewcommand{\arraystretch}{1}
            \centering
            \small
            \begin{tabular}{lccccc}
                \hline
                \textbf{Model} & \textbf{Unique} & \textbf{Relevant} & \textbf{\% Relevant} & \textbf{Contained in Nadi} & \textbf{New relevant} \\
                 & \textbf{papers} & \textbf{papers} & & \textbf{(of 16 possible)} & \\
                \hline
                ChatGPT      & 26 & 22 & 85\% & 14 & 8 \\
                ScholarAI             & 21 & 12 & 56\% & 7  & 5 \\
                ScholarAI GPT & 20 & 14 & 70\% & 7  & 7 \\
                \hline
                Combined (unique) & 40 & 26 & 65\% & 15 & 11 \\
                \hline
            \end{tabular}
            \caption{Performance of each GenAI model in the source-identification task.}
            \label{tab:source_summary}
        \end{table}

        We observed ScholarAI and ScholarAI GPT tended to identify more papers that were not formally published (e.g. theses and arXiv preprints). This is consistent with the model's design to primarily search within academic databases that contain a large catalogue of open-access articles.

        In terms of hallucinations, no fictional papers were identified. This is an improvement from past literature which identified high hallucination rates with past models of GPT. However, we noticed some incorrectly attributed authors from ChatGPT and broken links from ScholarAI GPT. Also, ScholarAI specifically recommended some papers which were not relevant and about a completely different topic.

        Furthermore, we attempted to use ChatGPT to organize the literature it found into our categories of relevant and not relevant. ScholarAI was not considered as it was unable to provide suitable responses to this task. In all scenarios, ChatGPT displayed a 20\% false positive rate and a 15-22\% false negative rate. In these false negatives, we noticed papers which were especially clear in their relation to dWOLS (i.e., containing "dWOLS" in the title or abstract), and yet the model did not classify them correctly.  

        Finally, we prompted ChatGPT to reproduce a summary table similar to one that appears in the Nadi review to categorize the relevant papers based on their contribution to dWOLS. We found ChatGPT was unable to produce sufficient output, as the resulting table contained duplicate entries, incorrect classification, and hallucinations for details such as existence of a real-data application.
            
    \subsubsection{Summarizing Sources}
        Next, we explored the use of GenAI in creating summaries for the sources it identified for the literature review. This application could assist a researcher in screening each source for relevancy, as well as extracting information to use in the literature review itself.
        
        In the previous section, we identified 40 unique papers. However, 2 of these papers were published before the original dWOLS paper, and another 2 papers were not in relation to dWOLS in any capacity, thus they were removed. We will also consider the 1 paper that appeared in the Nadi review but was not found by GenAI. Thus, we analyzed summaries for 37 papers in this section. We designed a standard prompt to use across all GenAI models to summarize each paper. 
         
To avoid GenAI from accessing a pre-print version of a paper, we will prompt twice: The first prompt will include the paper’s abstract, and request a summary based on the given abstract. The second prompt will require uploading a PDF file of the source paper, then requesting a summary of the paper. 
       
To encourage consistency in responses, particularly with the length of each summary, the full-text summary prompt will request a structure with specific subsections. 

We will focus on evaluating the summary results for ChatGPT and ScholarAI.

            \greybox{ 
            \textbf{Abstract-only summary prompt:}\\
            I am writing a literature review on DTR estimation using dWOLS methodology introduced in Doubly-robust dynamic treatment regimen estimation via weighted least squares by Wallace and Moodie. I am looking for extensions dWOLS methodology and real data analysis. For the following paper, use only the abstract and provide a summary for the literature review.
            \\Paper: (insert paper citation)
            \\Abstract: (insert paper abstract)}
            \vspace{-1em}
            \greybox{
            \textbf{PDF summary prompt:}\\
            I am writing a literature review on DTR estimation using dWOLS methodology introduced in Doubly-robust dynamic treatment regimen estimation via weighted least squares by Wallace and Moodie. I am looking for extensions dWOLS methodology and real data analysis. Summarize the following paper and include subsections of Main Idea and Contribution, dWOLS Extension, Methodology, Applications, and a short paragraph summary for the literature review. Use only the pdf of the paper: (insert paper citation/pdf file name).}

            The summaries created by GenAI had minimal apparent hallucinations and appeared true to the source material. We will proceed with commenting on the structure, quality, and content of the summaries. More details on each model's performance can be found in the Appendix section \ref{summ}.

            The summaries created by GenAI had minimal apparent hallucinations and appeared true to the source material. We will proceed with commenting on the structure, quality, and content of the summaries. More details on each model's performance can be found in the Appendix section \ref{summ}.

A major difference between the output of ScholarAI and ChatGPT was the structure of the abstract summaries. ScholarAI remained extremely consistent, containing the exact same subsections throughout each summary. However, ChatGPT varied considerably with the structure of the summaries it created, not using consistent subsections or headings. In particular, ScholarAI included general background information which made its summaries more suited for a reader who is not already very familiar with dWOLS and DTR estimation. For the full-text summaries, since the prompt instructed the model to include specific subsections, ChatGPT displayed improved consistency in structure.

We observed ScholarAI's summaries were more likely to omit critical details about the paper, particularly when summarizing the full-text (see the Appendix section \ref{summ_full}). ChatGPT, meanwhile, often provided more details, although still lacked in contextual understanding of the topic. Notably for the full-text summaries, ChatGPT's summaries occasionally included formulas and equations from the methodology of the paper, although without proper definition of variables or contextual information. As a result, these summaries may offer limited utility to readers who are not already familiar with the underlying paper or methodological and notational frameworks.
            
Both exhibited problems with uploading and reading PDF files. Repeated uploads were often necessary as the models struggled to locate and read the papers. ScholarAI had substantially more issues in sourcing from uploaded documents, even producing summaries when it was unable to read the PDF file and without notifying the user of this error (see Appendix section \ref{summ_full}). For example, one resulting summary was extremely vague and repeated the ideas from the abstract without further elaboration on the methodology of the paper. The summary states that the methodology employs ``advanced statistical methods" and ``the paper likely utilizes sophisticated modeling strategies." This could be classified as a form of (minor) hallucination, although the summary did not explicitly state untrue facts about the paper, rather being vague and lacking specific details. 
            
Regardless, the summaries from both models had questionable depth and contextual knowledge, rendering them of little use for a researcher who is unfamiliar with the content or wants to gain further understanding. Thus, these summaries are limited in scope, and cannot be simply copied into a potential literature review without substantial human oversight.

    \subsubsection{Identify Gaps in Literature}
        Identifying relevant research gaps is essential for literature review development. This step requires contextual knowledge of the topic at hand, including the current scope of research. While GenAI has demonstrated promising abilities in terms of literature retrieval and summarization, these tasks only test the model’s ability to search and reproduce information. Identifying gaps in research requires a critical review of the content.
        
 We prompted GenAI to identify gaps in the literature within the following contexts: 1. After requesting for abstract/PDF summaries; 2. After providing genAI with a list of citations; and 3. After providing genAI with uploaded full-text PDFs of the publications. These scenarios were chosen to replicate the workflow of a researcher, and to determine any advantages or disadvantages of certain contextual information available to the model. 

        When prompted to provide gaps in literature in the same sessions where the abstract and PDF summaries were created, both GenAI struggled to provide relevant results. Both models were unable to fully draw on the contextual memory of the session, rather providing gaps that were actually explored in the provided literature. For example, ChatGPT suggested extensions on the methodology to ordinal outcomes and interference beyond dyads, both of which were discussed in the literature the model summarized. 
        
        Providing context through a list of citations or uploaded papers improved the relevancy of the results. ChatGPT and ScholarAI provided a substantial number of gaps through repeated and varied prompting. In the gaps provided, most were an aggregation of the identified gaps that appeared in the literature already, were out-of-scope, or lacked understanding of the research domain. For example, ChatGPT suggested extensions in domains such as pediatrics, although DTR estimation is typically applied to chronic conditions. ScholarAI suggested combining dWOLS with causal inference techniques, although the dWOLS method was developed within a causal inference framework. 
        
        Comparing the two GenAI models, we observed that ChatGPT was able to provide more focused, actionable, and grounded suggestions than ScholarAI. ChatGPT suggested extensions to more outcome types such as counts, recurrent events, or patient-reported outcomes. ScholarAI did not provide any suggestions for extensions to different outcome variate types. Nonetheless, both models suggested a number of high-level gaps that would require a domain expert to review. A detailed comparison of the gaps provided by the models can be found in the Appendix section \ref{gap}.

    \subsubsection{Creating a Literature Review}

        We have discussed the ability of GenAI as a tool to aid specific tasks within a literature review, such as finding and summarizing sources. With the abilities of GenAI to write and synthesize information, we will now evaluate how GenAI performs when prompted to write an entire literature review.

        Although ScholarAI was advertised as being able to conduct literature reviews, no full written review could be produced even through many repetitions and variation of prompts. With ScholarAI unable to produce successful results, we focused on ChatGPT which has proven capabilities to write longer texts. We compared the GPT-5 Thinking and the GPT-5 ``Deep Research" model.
        
        As expected, both ChatGPT models generated substantial text which synthesized the provided literature, drew connections between the papers, and identified gaps in the research. However, within each literature review we identified problems with citations, structure, and lack of depth of understanding of the content. Full details about these attempts can be found in the Appendix section \ref{LR}.

      Both ChatGPT models exhibited major issues when attributing papers to their correct topics. In particular, both had instances where an incorrect paper was cited for a topic, or the topic was discussed without any citation provided. The Thinking model in particular struggled to provide a cohesive structure that would be expected of an effective literature review. Instead, the output included bullet points, many subheadings, no background information, and lack of detail and discussion of the papers. In comparison, Deep Research provided a literature review which was of better quality in terms of writing and structure, but struggled in accessing relevant papers. Specifically, this model is designed to search the internet instead of using the user's uploaded documents as a primary source. Thus, it only discussed open-access papers it was able to find, which does not cover the full scope of research we expected. As well, some oversights in terms of content were observed. While the Deep Research model included an introduction with background information about dWOLS for the literature review, it did not properly define any notation it introduced. It also struggled with basic contextual statistical assumptions - the model wrote that a distributional assumption is required for dWOLS as a least squares model - however, this is a common misunderstanding, as a distributional assumption is not required when estimating parameters in this context. This again displayed how ChatGPT lacks depth in understanding the topic.

\section{Generative AI Use in Statistical Methodology Code Development}
    
    We now turn to the use of GenAI to write code based on statistical methodology papers. For this, we will use ChatGPT version GPT-5 only, as ScholarAI does not exhibit programming capabilities. We will request ChatGPT provide R code, which also has minimal appearance in the current scope of research on GenAI use. 

    We will focus on coding tasks based on the dWOLS methodology and an extension: Generalized doubly-robust weighted ordinary least squares (G-dWOLS, \citeauthor{Schulz2021}, \citeyear{Schulz2021}), which extends dWOLS to accommodate continuous treatments (dosages).

    In particular, we will 1. Ask ChatGPT to reproduce simulation studies presented in the original dWOLS paper and the extension paper; 2. Ask ChatGPT to analyze simulated datasets using the original dWOLS method; and 3. Ask ChatGPT to extend \emph{DTRreg} \citep{Wallace2017}, an existing dWOLS R package to incorporate the G-dWOLS extension.

    \subsection{Background on dWOLS}

    We first provide a high-level overview of dWOLS methodology to provide context when discussing the code produced by ChatGPT. For simplicity, we consider a case where a single treatment decision is to be made, but note that the method can be applied to a case where multiple treatment decisions are made over a series of \emph{stages}.
    
    We consider a continuous outcome $y$ where, without loss of generality, larger values are preferred. A binary treatment $a \in \{0, 1\}$  is received, where $a = 0$ may be thought of as no treatment or a reference treatment (such as standard care). Lastly, we consider a single \emph{tailoring variate} $x$ which may interact with treatment, such as age or symptom severity.
    
    In the simplest case, we might propose an \emph{outcome} model
\[
E[Y|X = x, A = a] = \beta_0 + \beta_1 x + a(\psi_0 + \psi_1 x)
\]
where our goal is to identify the treatment $a \in \{0, 1\}$ that maximizes this expectation. Written this way, we can see that our choice depends only on the $\psi_0, \psi_1$ terms, with $a = 1$ being optimal if $\psi_0 + \psi_1 x > 0$, and $a = 0$ otherwise.

Framed this way, the outcome model is compartmentalized into two components: the \emph{treatment-free} model (parameterized by $\beta$) and the \emph{blip} model (parameterized by $\psi$). As such, the treatment-free model is a nuisance model with respect to our analytical goal of identifying the optimal treatment decision rule.
    
In practice, the treatment and tailoring variates may not be independent (such as if older patients are more likely to receive a particular treatment). Therefore if the treatment-free model is misspecified, the estimates of the blip parameters may be biased. dWOLS addresses this by fitting a third model - the \emph{treatment} model - which models the relationship between treatment and tailoring variates, such as via logistic regression.

dWOLS then carries out a weighted least squares regression, using weights derived from the treatment model. A family of weight functions may be used, with the paper using weights of the form $w = |a - P(A = a|X = x)|$: the absolute difference between the observed treatment and the probability it was received. The blip parameters are then consistently estimated if at least one of the treatment-free or treatment models is correctly specified (along with the blip model), a property known as \emph{double robustness}. dWOLS can be extended to the case of multiple treatment decisions over time, using a backwards induction technique that begins by estimating the final treatment rule and then estimating each of the earlier treatment rules in a stepwise manner.

\subsection{Results}
    \subsubsection{Writing R code based on academic papers}

    We first evaluate the use of ChatGPT as a coding assistant for implementing dWOLS and extensions. dWOLS and G-dWOLS both have published code to implement their methods, which we used as a point of comparison for the results from ChatGPT. We presented various scenarios and contextual information to ChatGPT to replicate different research situations. For example, we varied whether the source paper was uploaded and whether or not we supplied our own code or specified certain models (such as the treatment, treatment-free, and blip models).

    When evaluating the code, we considered the following questions: Did the code execute? Did it use the correct methodology? Did the code produce the expected parameter estimates? Details about prompts and results from specific scenarios can be found in the Appendix section \ref{code}. The following is an example of a prompt used for this task:

    \greybox{\textbf{Prompt:} Write R code that implements the dWOLS function from the uploaded paper for the following code that provides the data and models: (insert R code with models)}

   When we did not upload the paper to ChatGPT, it searched the internet and suggested to use the R package DTRreg, or it attempted to create a function that was not sourced from the original paper methodology. 

   When provided with the source paper, ChatGPT was able to produce code, although many issues were observed. In almost every session, the first code blocks produced by ChatGPT was not executable. The code always included a small error, such as incorrect syntax, or calling a variable that was not defined. When these errors occurred, we explained the error to ChatGPT in a response and it was usually able to fix and produce code that executed in an additional 1-2 prompts. This was not always the case; in a few situations we were unable to elicit executable code from ChatGPT through this conversational approach.
    
    When executable code was produced, further problems were observed. When producing code using the original dWOLS paper, ChatGPT excluded a parameter it was expected to estimate for reasons that could not be determined. When this error was fixed and ChatGPT produced the correct set of estimates, we found that the results were only slightly different from the expected results. 
    
    This difference was due to ChatGPT using incorrect methodology, in which it used incorrect estimating equations and different weight functions. Notably, ChatGPT cited the reason for its different approach was to preserve the double-robustness of the method, but this is not backed up by the source paper. Attempts to resolve these issues through conversing with the model were limited by the model’s tendency to agree with the user suggestions, even when the proposed methodology was incorrect. In this context ChatGPT was not a reliable source to understand and program the methodology of the paper, as it appears to have hallucinated a method that does not exist in the provided source material.  

    When programming the G-dWOLS method, prompting needed to be more specific as this paper included more options based on the research setting. For instance, the original dWOLS paper uses one specific weight function within a family of weights, while the G-dWOLS paper proposes several weight functions one could utilize. When not specifying what weights to use or number of stages, ChatGPT assumed a single-stage setting and the simplest weights suggested in the paper. This left little flexibility, and further prompting was required to generate code for more complex scenarios. 
    
    Again, we observed some small errors which impacted the results of the method. These included an incorrect number of parameter estimates, slightly different estimates due to an alternative method in calculating the weights, and output which was reversed (e.g. estimates for the second treatment decision were labeled as for the first treatment decision, and vice-versa). When these small differences were fixed, we once more observed a hallucination regarding the methodology of the paper. ChatGPT used incorrect estimating equations, featuring parameters which should not have been included in the models. Both these small and large errors are worrisome as they only altered the parameter estimate results slightly, and might not be obvious without the required background knowledge of these methods and careful scrutiny of the code.

    While ChatGPT’s code exhibited many errors and incorrect understanding of the underlying methodology, we found that there were some positives to the model’s performance. ChatGPT was quick in generating functional code, only requiring 1-3 prompts. The code rarely made errors regarding the algorithmic nature of the methods, for instance, the model components and backward induction steps. The main issues we observed arose in the form of the estimating equations and weight functions used. After editing ChatGPT's code to use the correct estimating equations and weights, the code produced correct results, which shows that the underlying algorithm was functional.

    \subsubsection{Writing R code and applying dWOLS to simulated data}

    We now evaluate ChatGPT’s ability to apply the dWOLS method when only provided a dataset. This attempts to simulate a researcher using GenAI to carry out a real data analysis without directly writing their own code. We simulated two datasets and presented them to ChatGPT as if they were real data. One dataset was considered `simple', with the other more complex, with complexity varying in the models used in the data generation process. The code used to create these simulations and prompts used can be found in the Appendix section \ref{sim}. The context provided to the model included varying whether the dWOLS paper was uploaded, and including code that specified the form of the treatment, treatment-free, and blip models we wished to use.

    Example prompt: 
    \greybox{\textbf{Prompt:} I have uploaded health\_data.csv which contains patient information for two stages of a DTR. In stage 1, we observed patient characteristic Risk1, prescribed Treatment1, and in stage 2 we observed patient characteristic Risk2, and prescribed Treatment 2. Treatment is a drug, where 1 means we give the patient the drug, and 0 means we give the patient a placebo. Finally, Health is the final measure of health for a patient, where higher values are better. Use the uploaded paper to write R code estimate the optimal DTR using dWOLS for this data.}
    
    Without uploading the dWOLS paper, ChatGPT suggested to use the R package DTRreg, and gave example code to run it on the provided dataset. However, it hallucinated a function that does not exist in the DTRreg package. Upon further prompts, ChatGPT agreed that the function did not exist, and opted to create its own dWOLS function. Given the specification of the models, ChatGPT successfully used the DTRreg package, and provided correct estimates. In this case, ChatGPT was able to source the DTRreg package information from the internet and use the correct syntax.

    When coding the dWOLS function from scratch, ChatGPT produced the same errors observed in the previous section, including incorrect estimating equations and weights. These findings were consistent across the simple and complex simulations, and when not specifying the models, ChatGPT always defaulted to the simple model structures used as examples in the dWOLS paper. Although we prompted the model to consider more complex model structures, attempts to implement these alternatives resulted in more errors that could not be resolved through continued prompting.

    \subsubsection{Extending the R package}

    Finally, we will explore if GenAI can be used to extend the R package DTRreg to incorporate G-dWOLS. We uploaded the G-dWOLS paper to ChatGPT, and asked a variety of prompts to attempt to extend the DTRreg package. We evaluated how the model responded when we provided the DTRreg package code or source files. As well, we focused on conversing with the model and followed its instructions to update the R package. When the instructions did not work, we prompted the model with the error message, and followed the new instructions it provided.

    The prompts can be found in Appendix section \ref{extendR}. An example prompt is given:

    \greybox{\textbf{Prompt:} Using the attached paper, extend the DTRreg function in the DTRreg R package to implement G-dWOLS. Include an example to run the function.}

    When we prompted ChatGPT to extend the DTRreg package to include the G-dWOLS method but did not provide any package-specific context, ChatGPT created a G-dWOLS function with guidance for how it can be implemented in the package. However, these integration instructions were often inconsistent with the actual DTRreg code. For example, it recommended we modify the existing ``DTRreg" function, but the code blocks it referenced did not exist in the actual source code. In further prompts, ChatGPT was able to access the source code of the DTRreg function, however, the suggested modifications did not succeed in implementing the method. 
    
    When providing ChatGPT with the source files for the package, the model attempted to guide the integration process via Git-based edits. However, this approach was not successful, either due to incorrect instructions or incorrect code in the patch file. The model then recommended applying the changes manually, but the resulting implementation still did not correctly implement the G-dWOLS method, seen through the incorrect number of parameter estimates produced. We were therefore unable to use ChatGPT to develop code that could implement the G-dWOLS method and update the package effectively.

\section{Discussion}

\subsection{Literature Review Task}

    This case study was motivated by the rapid adoption of GenAI across many domains and industries. Due to its strong capabilities in text generation and information synthesis, GenAI has potential applications in the research process. We sought to explore current risks and limitations of using GenAI within research, as understanding its drawbacks is crucial to using these tools effectively. We focused on applications within statistical methodology and epidemiological research, which has received little previous attention. We explored GenAI use in the creation of a literature review in the context of DTR estimation via dWOLS. Our results suggest the current state of GenAI is not an adequate replacement for a human researcher’s work, particularly regarding domain expertise and critical thinking. Table~\ref{tab:scorecard} summarizes the relative performance of each GenAI model across the major stages of the literature-review workflow, together with the dominant failure modes identified for each task. Table~\ref{tab:recommendations} summarizes recommended uses of GenAI and the level of human oversight required.

     \begin{table}[H]
         \centering
         \small
        \caption{Task-by-task evaluation of the three GenAI models for creating a literature review. \Checkmark\Checkmark\ = comparatively strong performance; useful as a starting point but requires expert verification. \Checkmark\ = limited/moderate utility; requires substantial expert verification and revision. \textbf{--}\ = unable to produce usable output despite repeated prompting.}
         \renewcommand{\arraystretch}{1.2}
         \begin{tabular}{p{2.5cm}cccp{5cm}}
             \hline
             \textbf{Task} & \textbf{ChatGPT} & \textbf{ScholarAI} & \textbf{ScholarAI GPT} & \textbf{Main failure mode} \\
             \hline
             Identify sources                & \Checkmark\Checkmark & \Checkmark            & \Checkmark            & Missed relevant papers lacking explicit ``dWOLS'' terminology; inclusion of irrelevant or weakly related papers \\
             Determine \newline relevancy             & \Checkmark            & \textbf{--}           & Not evaluated        & 15--22\% false-negative rate, including papers explicitly containing ``dWOLS'' in the title \\
             Summarize sources               & \Checkmark\Checkmark & \Checkmark            & Not evaluated           & Limited depth and contextual understanding; PDF read failures produced vague or unsupported summaries \\
             Identify \newline research gaps          & \Checkmark            & \Checkmark            & Not evaluated          & Aggregated previously identified gaps rather than generating novel insights; limited domain-specific understanding \\
             Write full \newline literature review    & \Checkmark            & \textbf{--}           & Not evaluated           & Produced coherent text but lacked reliable citation attribution and sufficient methodological depth\\
             \hline
         \end{tabular}
         \label{tab:scorecard}
         
     \end{table}

\begin{table}[H]
    \centering
    \small
    \renewcommand{\arraystretch}{1.25}
\caption{Task-by-task recommendations for researchers using GenAI to support a statistical literature review.}
    \begin{tabular}{p{3.0cm} p{5.5cm} p{5.5cm}}
        \hline
        \textbf{Task} & \textbf{Recommended use of GenAI} & \textbf{Required human oversight} \\
        \hline
        Source identification and relevancy screening
        & Use multiple models with complementary search domains; vary prompt wording and repeat prompts across sessions to widen coverage. GenAI may also assist with first-pass relevancy screening when citation lists or PDFs are supplied.
        & Verify every citation against the original source and re-screen papers classified as ``not relevant''; expect missed papers lacking explicit topic keywords and false-negative rates of approximately 15--22\%.
        \\
        Summarize sources
        & Request a fixed subsection structure to improve cross-comparability; use summaries for rapid triage and high-level familiarization before expert review.
        & Verify key technical and methodological claims against the original paper; treat vague or generic summaries as a possible indication of PDF read failure or limited contextual understanding.
        \\
        Identify \newline research gaps
        & Use GenAI to aggregate and organize research gaps identified across multiple papers; repeated prompting may help surface different perspectives or themes.
        & A domain expert must distinguish genuinely novel directions from generic, redundant, or out-of-scope suggestions before pursuing proposed gaps.
        \\
        Write full \newline literature review
        & Limit use to paragraph-level drafting assistance (e.g., background text or transitions) rather than full-document generation.
        & Re-check every citation--claim pairing; expect misattributions, incomplete literature coverage, and silent omission of important methodological details or assumptions.
        \\
        \hline
    \end{tabular}
    \label{tab:recommendations}
\end{table}

Some results were encouraging. Despite GenAI historically producing hallucinations, ChatGPT and ScholarAI identified relevant literature without providing entirely fictitious papers. However, there were substantial details in the provided citations which were incorrect, and a number of false positives (papers identified as relevant which, upon review, were not) which solidifies the need for manual verification. 

Repeated prompting with varied wording improved search breadth, and utilizing multiple models with different search domains was beneficial as each model identified relevant papers that were missed by the others. Additionally, we found that both models exhibited issues and inaccurate results with determining relevancy of a given paper to the literature review, even incorrectly dismissing papers that were entirely explicit in their relation to dWOLS (i.e., including keywords such as "dWOLS" in the title or abstract). 

ChatGPT and ScholarAI performed moderately well in summarizing the literature, as reflected by the lack of fictitious papers or incorrect information in any of the summaries. However, we identified limitations in the quality, content, and structure. Without specific prompting, ChatGPT was not consistent in length, content, or structure of each summary, making it difficult for cross-comparison between papers. Structured prompting was required for ChatGPT to provide a consistent structure. On the other hand, ScholarAI was entirely consistent in structure and content through multiple prompting strategies. While this allowed for easier comparison with other summaries, ScholarAI was unable to identify specific details that might be of interest to a researcher. 

In general, the summaries produced by GenAI were able to produce the main idea of the text, but lacked depth and contextual understanding of the topic. This results in an unreliable level of quality in each summary and requires an expert researcher to review the paper independently. Additionally, both models displayed issues with uploading and reading PDF files. There were cases where ScholarAI was unable to read the paper and provided entirely vague summaries instead of notifying the user that it was unable to source any information from the document.

When identifying research gaps, GenAI primarily reproduced gaps already identified in the relevant literature, rather than new ideas. Results improved when providing the models clear contextual information such as a list of relevant literature. Prompting multiple times also improved results as the models occasionally highlighting new themes, although, many gaps provided were high-level or lacked contextual understanding. A review from a domain-expert is necessary to verify the quality of the generated gaps. 

We concluded by attempting to produce an entire literature review using ChatGPT, as ScholarAI was unable to generate sufficient content for a literature review. Although ChatGPT was able to produce longer formatted draft which summarized many of the relevant papers, issues were identified in citations, content, and overall quality. ChatGPT repeatedly attributed papers incorrectly or omitted references altogether. We also noticed some minor content errors that could be attributed to the model’s lack of contextual knowledge for the topic. More specific prompting may make GenAI useful for drafting smaller sections, but substantial human supervision is still required. Currently, generating a complete, high-level literature review without significant human oversight is beyond the capabilities of these existing GenAI models. 

\subsection{Coding Task}

Finally, we explored the use of GenAI in coding-based tasks. While the use of GenAI as a coding assistant is well documented, we attempted to simulate a researcher attempting various tasks pertaining to the implementation of statistical methods based on their associated papers. While we found that in some specific circumstances it was possible to produce accurate code, it always required multiple prompts, with most initial attempts resulting in code that could not even be executed. What is perhaps most concerning, however, was the observed capacity for GenAI to produce code that would run and produce results that were close to, but not exactly, correct. This highlights a major concern in the use of GenAI as a coding assistant, where incorrect code could be produced that appears functional, with its errors only identifiable following careful scrutiny from an expert. As such, our case study suggests considerable care must be exercised if using generative AI tools in coding statistical methodology. Table \ref{tab:scorecard_code} summarizes the relative performance of each GenAI model across the code generation tasks, along with the main failure identified in each task. Table \ref{tab:recommendations_code} summarizies recommended uses of GenAI for code generation and details regarding the level of human oversight required.

   \begin{table}[H]
         \centering
         \small
        \caption{Task-by-task evaluation of the three GenAI models for code generation. \Checkmark\Checkmark\ = comparatively strong performance; useful as a starting point but requires expert verification. \Checkmark\ = limited/moderate utility; requires substantial expert verification and revision. \textbf{--}\ = unable to produce usable output despite repeated prompting.}
         \renewcommand{\arraystretch}{1.2}
         \begin{tabular}{p{2.5cm}cccp{5cm}}
             \hline
             \textbf{Task} & \textbf{ChatGPT} & \textbf{ScholarAI} & \textbf{ScholarAI GPT} & \textbf{Main failure mode} \\
             \hline

             Write R code \newline based on academic papers & \Checkmark            & Not evaluated            & Not evaluated           & Consistently hallucinated methodological details and defaulted to simple model specifications\\

            Write R code to apply dWOLS to simulated data    & \Checkmark            & Not evaluated            & Not evaluated           & Consistently hallucinated methodological details and defaulted to simple model specifications\\
             
             Extend R \newline package  & \textbf{--}            &  Not evaluated           & Not evaluated           & Unable to modify existing R package; produced incorrect integration instructions and code modifications \\
             \hline
         \end{tabular}
         \label{tab:scorecard_code}
         
     \end{table}
     
\begin{table}[H]
    \centering
    \small
    \renewcommand{\arraystretch}{1.25}
\caption{Task-by-task recommendations for researchers using GenAI to support code generation.}
    \begin{tabular}{p{3.0cm} p{5.5cm} p{5.5cm}}
        \hline
        \textbf{Task} & \textbf{Recommended use of GenAI} & \textbf{Required human oversight} \\
        \hline
        Write R code \newline based on academic papers
        & Use as a coding assistant to rapidly generate simple implementations of methodological descriptions into code.
        & Every detail of the methodology should be carefully reviewed and validated to ensure it matches the source paper
        \\
        Write R code to apply dWOLS to simulated data
        & GenAI can assist with data preparation and utilizing the R package when model specifications are clearly defined.
        & Verify correct model specification, confirm package functions are used properly, and validate implementation against known or simulated truth.
        \\
        Extend R package
        & Use GenAI to explore implementation ideas and as a coding assistant to draft basic code modifications.
        & Manually review all proposed code changes and test thoroughly, ensuring correct methodology.
        \\
        \hline
    \end{tabular}
    \label{tab:recommendations_code}
\end{table}

\subsection{Limitations and Future Research}

This case study is not without limitations. The online existence of the Nadi review we attempted to replicate might have influenced the model’s responses. ChatGPT and ScholarAI are inherently not transparent with its reasoning methods and sources, so we are unable to be completely confident that this resource was not accessed.
    
Due to the nature of Generative AI, no two responses are the same. Thus, while we employed prompting techniques to gain a sense of how much responses changed between sessions, it is possible that there is unaccounted for variation that would affect our results. For instance, while we did not find major hallucinations when searching for sources or summarizing, it is still possible that hallucinations may occur with different prompts or contextual information. This lack of consistency makes it difficult to assess the performance of GenAI more generally. 
    
This case study took an exploratory approach when evaluating output from GenAI, but further research could explore more formal approaches. For example, summaries could be assessed using a rubric, graded based on quality, structure, the number of hallucinations, and so on. Another GenAI model could be tasked to automate this grading process.
    
In this case study, we employed the use of zero-shot prompting, in which no examples were provided to the model in the initial prompt. Other prompting methods, such as few-shot prompting (which includes an example), could be used and may change the model responses. 

Lastly, while this case study focused on the used of ScholarAI and GPT-5, these technologies update extremely quickly. In the span of completing this case study, ScholarAI was acquired by Jenni AI and OpenAI released new models GPT-5.1 and GPT-5.2. The evolutionary aspect of these models presents difficulties in researching the performance of GenAI, as new updates can increase or decrease performance in a variety of tasks.

\subsection{Conclusion}
    
    Although we do not recommend using GenAI without careful guidance, small adaptations of this technology have the potential to increase the efficiency of a researcher. For example, GenAI generated summaries that provided a useful high-level overview of each paper while also capturing details at the user's request that might be absent from the abstract. As well, we saw considerable improvement from past published findings regarding hallucination rates of ChatGPT’s ability to gather literature sources. Using GenAI to source literature takes advantage of the model’s advanced capabilities to search the internet but one must still diligently review each source found, as the model can lack accuracy. Thus, any current adaptation of GenAI tools in the literature review process or programming must be approached as a secondary tool to a researcher’s own knowledge.

    We advocate that future scholarly work that uses GenAI should be completely transparent and responsible, especially given well-documented concerns regarding bias and hallucinations. Additionally, we do not recommend GenAI should replace the role of the human researcher. Rather, it should be used as a tool to increase efficiency, effectiveness, and to streamline organizational and routine tasks. An optimistic use for this technology lies in making research more accessible. Given current limitations, GenAI is not a substitute for the expertise and intellectual originality of a human researcher. Significant work is still required to fully adapt these models for be reliable usage in academic research.

\newpage

\bibliographystyle{apacite}
\bibliography{bib1.bib}

\newpage
\setcounter{page}{1}
\section{Web Appendix} \label{appendix}

\subsection{Identify Papers - Prompts} \label{sources}
First, we present the list of prompts and follow-up prompts given to GenAI to identify relevant sources for the dWOLS literature review. The following sections contain details on each GenAI model's performance in finding sources. 

\vspace{0.3em}
\textbf{Prompts: }
\vspace{-0.3em}
\begin{enumerate}
    \itemsep0em
    \item Find articles that incorporate and extend the methodology from ``Doubly-robust
dynamic treatment regimen estimation via weighted least squares” by Wallace \& Moodie.
    \begin{enumerate}
        \itemsep0em
        \item Find more articles that incorporate and extend the dWOLS methodology from
    ``Doubly-robust dynamic treatment regimen estimation via weighted least squares” by Wallace \&
    Moodie and provide title, date published, author names, and abstract.
    \end{enumerate}
    \item Find articles that use the dWOLS methodology in real data analysis from ”Doubly-
robust dynamic treatment regimen estimation via weighted least squares” by Wallace \& Moodie.
    \begin{enumerate}
        \itemsep0em
        \item Provide more articles from any time period that have not already been mentioned.
    \end{enumerate}
    \item Find articles that incorporate the dWOLS methodology from ``Doubly-robust dynamic treatment regimen estimation via weighted least squares" by Wallace \& Moodie and provide abstracts. I want articles that 1. Use/extend dWOLs methodology and 2. Apply dWOLS in real data analysis.
    \begin{enumerate}
        \itemsep0em
        \item Find even more articles that have not already been found. 
        \item Find every article you can that applies and/or extends dWOLs method that has not already been found
    \end{enumerate}
    \item Find 22 articles that incorporate the dWOLS methodology from ``Doubly-robust dynamic treatment regimen estimation via weighted least squares" by Wallace \& Moodie. I want articles that use/extend dWOLs methodology and apply dWOLS in real data analysis.
    \item Find articles to complete a literature review based on the dWOLS methodology from ``Doubly-robust dynamic treatment regimen estimation via weighted least squares" by Wallace \& Moodie. I want articles that use/extend dWOLs methodology and apply dWOLS in real data analysis.
    \item For the following prompts, do not read or consider the following paper in your answer: Recent advances in doubly-robust weighted ordinary least squares techniques for dynamic treatment regime estimation, Adel Ahmadi Nadi, Michael Wallace, https://www.arxiv.org/abs/2501.18819
    \begin{enumerate}
        \itemsep0em
        \item Find 20 articles that incorporate the dWOLS methodology from ``Doubly-robust dynamic treatment regimen estimation via weighted least squares" by Wallace \& Moodie. I want articles that use/extend dWOLs methodology and apply dWOLS in real data analysis.
    \end{enumerate}
    \item For the following prompts, do not read or consider the following paper in your answer: Recent advances in doubly-robust weighted ordinary least squares techniques for dynamic treatment regime estimation, Adel Ahmadi Nadi, Michael Wallace
    \begin{enumerate}
        \itemsep0em
        \item Find articles to complete a literature review based for DTR estimation using doubly-robust weighted ordinary least squares. I want articles that use/extend dWOLs methodology and apply dWOLS in real data analysis.
        \item Find me papers that discuss handling error in estimation for DTR with dWOLS method
    \end{enumerate}
    \item Make me a literature review on the topic of DTR estimation using dWOLS and provide 20 sources.
    \item Find me papers that discuss handling error in DTR estimation with dWOLS methodology or extensions of dWOLS methodology 
\end{enumerate}

            \subsubsection{ChatGPT}
            ChatGPT exhibited a large variety in the number of papers it provided, ranging from 3-12 on any given request. Typically, on subsequent prompts to find more papers, ChatGPT only provided 3-4 more sources, often including duplicates. These duplicates even occured when specifying to only provide articles that have already not been found.
            
            We attempted to prompt ChatGPT to provide a specific number of papers (e.g. Prompt 4). In some sessions, ChatGPT presented the specified number of papers on the first try. In other sessions, ChatGPT presented less than the specified amount, waited for user input to continue. The target number of papers was chosen (22) because at that point, ChatGPT had produced 22 unique papers and we wanted to evaluate if the model was about to reproduce these unique papers. In both cases, ChatGPT was unable to provide exactly 22 unique papers, rather presenting the same paper multiple times but citing two different aspects of its contribution (e.g. the paper included an extension and real data analysis). 
            
            No hallucinations of entire fictional papers were observed in any of the papers that were found by ChatGPT, which was an identified problem in previous versions of GPT. However, we observed some inconsistencies with the provided citations, for example, many papers were cited with incorrect authors. In subsequent prompts, the exact same papers continued to have issues with the author citation. Response Excerpt 1.1 and 1.2 includes a citation for the paper ``Variable selection in regression-based estimation of dynamic treatment regimes" by Zeyu Bian, Erica E. M. Moodie, Susan M. Shortreed, and Sahir Bhatnagar. In both of these responses, ChatGPT incorrectly credited Zhu and Laber. These authors appear as references in the paper but should not have been credited as authors.

            \greybox{\textbf{Response Excerpt 1.1 (ChatGPT):} 
            
            Variable selection within dWOLS: \\ Bian, Zhu \& Laber (2021). Discusses variable selection for regression-based DTR estimation and explicitly follows dWOLS, providing tools to regularize/choose tailoring variables.
            
            \textbf{Response Excerpt 1.2 (ChatGPT):} \\ Variable selection in regression-based estimation of dynamic treatment regimes (Biostatistics, 2021/2023 online) — Zhiqiang Bian; Ruoqing Zhu; Eric B. Laber. Real data: STARD depression dataset. Method: penalized dWOLS for variable selection; empirical illustration on STARD. }

            In some responses, ChatGPT cited the Nadi review, and appeared to rely on it to source relevant papers. To assess ChatGPT's performance without this dependency, we instructed ChatGPT to not read or reference the Nadi literature review. With this constraint, we continued to retrieve the same articles discovered in previous attempts. However, some subsequent responses referenced the Nadi review again, despite the explicit instruction. While ChatGPT sourcing from the Nadi review was worrisome, we still observed numerous articles which ChatGPT recommended that were not relevant for the literature review, and thus did not appear in the Nadi review. Note that the not relevant papers ChatGPT sourced were often of the topic of G-estimation or still related to DTR estimation, but not specific to dWOLS. Thus, we determined that ChatGPT did not solely depend on the Nadi review to source publications, as it still conducted an independent search.  

\begin{figure}[H]
                \centering
                \includegraphics[width=0.8\linewidth]{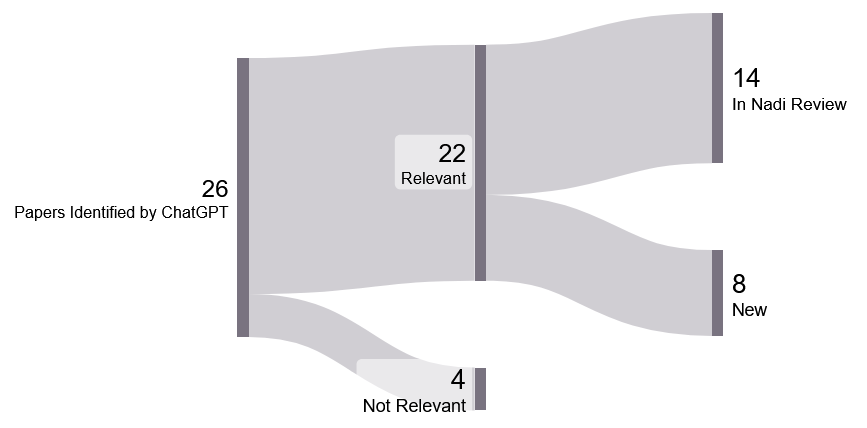}
                \caption{A Sankey diagram which shows the distribution of papers collected from ChatGPT. }
                \label{fig:sankey_chatgpt}
            \end{figure}

            ChatGPT retrieved 26 unique papers (Figure \ref{fig:sankey_chatgpt}). Only one of these papers were not officially published, rather appearing as a preprint on arXiv. Among the 26 papers, 22 were deemed relevant to the literature review topic (85\%). 14 papers that were identified were also found in the Nadi review, out of a total possible 16. Consequently, ChatGPT identified 8 new relevant sources.

        \subsubsection{ScholarAI}

            ScholarAI displayed difficulty in providing responses to the same prompts used with ChatGPT. In some cases, it reported issues with searching for papers, particularly when we prompted for “more” papers (Response Excerpt 2.1). ScholarAI exhibited less flexibility in terms of open-ended and less specfic prompts than ChatGPT. 

            \greybox{\textbf{Response Excerpt 2.1 (ScholarAI):} \\ To find multiple papers that use the ``Doubly-Robust Dynamic Treatment Regimen Estimation Via Weighted Least Squares" method, I'll conduct a literature search focusing on studies employing or referencing this technique. This search will also attempt to identify papers discussing related methodologies or applications in personalized medicine and dynamic treatment regimens. Let's proceed with finding these papers.  \\ Unfortunately, I wasn't able to retrieve any results directly related to the usage or applications of the ``Doubly-Robust Dynamic Treatment Regimen Estimation Via Weighted Least Squares" method using the initial approach.}

            When the model was able to retrieve papers, it often found papers entirely irrelevant to the subject, or were published prior to the target dWOLS paper For example, Response Excerpt 2.2 cites a paper which does not cite Wallace \& Moodie. The paper also isn't open-access, so ScholarAI is unable to read the paper, thus, this appears to be a hallucination. No other hallucinations were observed in terms of fictional papers or incorrect citations.

            \greybox{\textbf{Response Excerpt 2.2 (ScholarAI):} \\ Citations of Wallace \& Moodie's Work: A number of articles have cited and extended Wallace \& Moodie's methodological framework, particularly in adaptive enrichment designs for clinical trials. Some of these designs focus on biomarker-guided treatment and enriched patient populations in oncology trials.  ``An improved biomarker-guided adaptive patient enrichment design for oncology trials" (2025).}

            ScholarAI responded with a substantially lower number of sources per prompt, usually 3-5 papers, with the Nadi review included. If the number of papers was specifically requested, it often presented duplicates or papers that were entirely irrelevant to the subject. ScholarAI produced many sources that were not published in an academic journal, rather presenting thesis papers or pre-prints of publications. ScholarAI seems to prefer to source papers which are publicly available, which aligns with the model's capability to search sources like arXiv. Notably, ScholarAI presented relevant thesis papers which were not found by ChatGPT.
                        
\begin{figure}[H]
                \centering
                \includegraphics[width=0.8\linewidth]{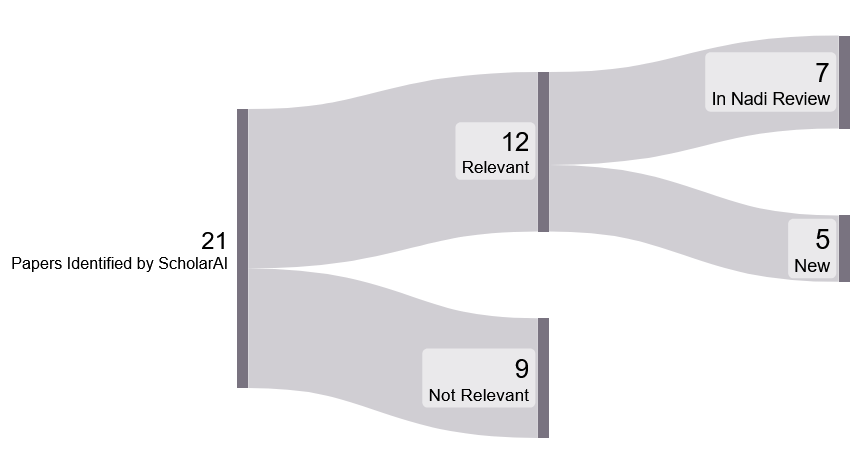}
                \caption{A Sankey diagram which shows the distribution of papers collected from ScholarAI.}
                \label{fig:sankey_scholarAI}
            \end{figure}

            ScholarAI provided 21 unique sources (Figure \ref{fig:sankey_scholarAI}), with 4 being unpublished thesis papers, and one was a book chapter. 12 out of the 21 papers were determined to be relevant (56\%). ScholarAI was able to identify 7/16 papers which appear in the Nadi review, and the model identified 5 new relevant papers.  

        \subsubsection{ScholarAI GPT}

            ScholarAI GPT exhibited issues as the model repeatedly reported errors that prevented it from retrieving papers (Response Excerpt 3.1) and required numerous repetition of prompts to gain relevant results.

            \greybox{\textbf{Response Excerpt 3.1 (ScholarAI GPT): } 
            \\Error talking to connector \\ It looks like my last search for additional dWOLS real-data applications not yet mentioned ran into a retrieval error. The query didn’t return new results this time.}

            On average, ScholarAI GPT presented 5-10 papers when unprompted for a specific target amount. As observed with other models, when prompted for a specific number of sources, it presented duplicates to reach the specified number. Notably, this model was able to discover articles that were not already found by ChatGPT and ScholarAI. No hallucinations were observed in terms of fictional papers or incorrect citations, however, the links it provided to the source’s journal page was often incorrect.
            
\begin{figure}[H]
                \centering
                \includegraphics[width=0.8\linewidth]{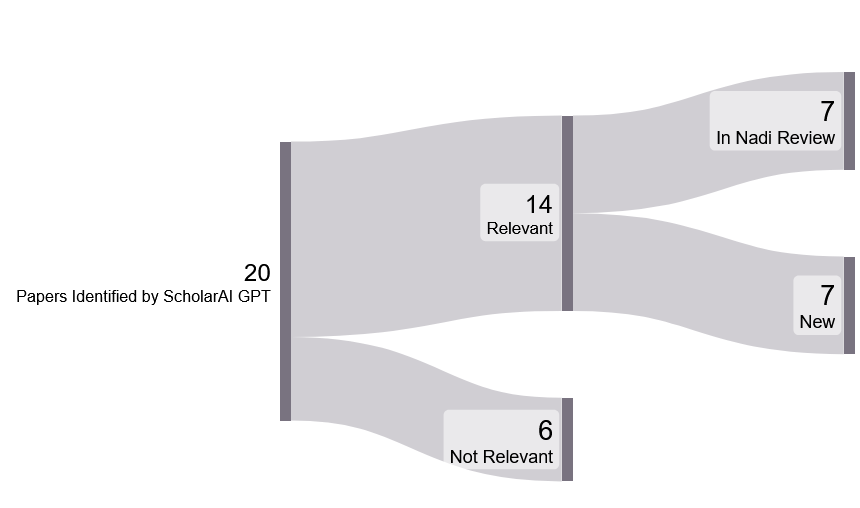}
                \caption{A Sankey diagram which shows the distribution of papers collected from ScholarAI GPT.}
                \label{fig:sankey_scholarAIGPT}
            \end{figure}

            ScholarAI GPT sourced 20 unique papers (Figure \ref{fig:sankey_scholarAIGPT}). Among these 20, 3 of them were unpublished thesis papers. Additionally, 14 out of 20 papers were deemed relevant (70\%). ScholarAI GPT was able to identify 7/16 papers which appear in the Nadi review, and the model identified 7 new relevant papers. 

        \subsubsection{Identify Relevancy}

        This section covers our attempt at using GenAI to identify the relevancy of given papers for the dWOLS literature review.

        \textbf{After abstract summaries/pdf summaries:} Based on the abstracts and pdf summaries, we requested GenAI to organize each paper into sections for the literature review:
            
            \greybox{\textbf{Prompt:} I am writing a literature review on DTR estimation using dWOLS methodology introduced in Doubly-robust dynamic treatment regimen estimation via weighted least squares by Wallace and Moodie. I am looking for extensions dWOLS methodology and real data analysis. Organize the previous papers into the following categories: Extensions of dWOLS, Assessing/Extending dWOLS assumptions, Not relevant to literature review. }
            
            Both GenAI were unable to organize every paper that was previously mentioned in the session, which might provide evidence that we reached a limit to the capacity of contextual knowledge in a single session. For the papers that were organized, we observed many errors, particularly with papers that appeared in the Nadi review being classified as not relevant.
                
        \textbf{List of citations:} For the next scenarios, only ChatGPT will be evaluated because ScholarAI was consistently unable to organized our entire list of citations, only opting to discuss 3-5 papers from the list. In the following prompt, we included all citations in APA-style:
            
            \greybox{\textbf{Prompt:} I am writing a literature review on DTR estimation using dWOLS methodology introduced in Doubly-robust dynamic treatment regimen estimation via weighted least squares by Wallace and Moodie. I am looking for extensions dWOLS methodology and real data analysis. Organize my list of citations into the following categories: Extensions of dWOLS, Assessing/Extending dWOLS assumptions, Not relevant to literature review. (Insert list of citations)}

            In multiple responses, a large portion of the papers were categorized the same each time. The most variation between each response was a small number of papers switching between the first and second categories. The papers that were deemed “not relevant” were identical in each session. There were only two papers are not relevant that were included incorrectly, and these papers were the same each time. These papers were related to DTR estimation but concerned G-estimation instead of dWOLS, and notably, shared the same author as the original dWOLS paper. Table \ref{tab:table_R} displays the confusion matrix for the classification of papers by ChatGPT.

            \begin{table}[H]
                \centering
                \begin{tabular}{|c|c|c|}
                \hline
                     & ChatGPT Relevant & ChatGPT Not Relevant \\
                     \hline
                 Relevant & 21 & 6\\
                    \hline
                 Not Relevant & 2 & 8\\
                    \hline
                \end{tabular}
                \caption{Confusion matrix for the classification of papers by ChatGPT using a list of citations}
                \label{tab:table_R}
            \end{table}

            \begin{itemize}
            \itemsep0em
                \item \textbf{True positive rate:} the probability that GenAI classified the paper as relevant (in the first two categories) and it actually is relevant, is 80\%
                \item \textbf{True negative rate:} the probability that GenAI classified the paper as not relevant and it actually is not relevant, is 80\%
                \item \textbf{False positive rate:} the probability that GenAI classified the paper as relevant when it actually is not relevant is 20\%
                \item \textbf{False negative rate:} the probability that GenAI classified the paper as not relevant when it actually is relevant is 22\%
            \end{itemize}

         \textbf{Uploaded papers:}
         The following prompt was utilized in ChatGPT to organize the uploaded articles into categories for the literature review:
         
            \greybox{\textbf{Prompt:} I am writing a literature review on DTR estimation using dWOLS methodology introduced in Doubly-robust dynamic treatment regimen estimation via weighted least squares by Wallace and Moodie. I am looking for extensions dWOLS methodology and real data analysis. Organize the uploaded pdfs into the following categories: Extensions of dWOLS, Assessing/Extending dWOLS assumptions, Not relevant to literature review.}

            Again, through multiple responses, a large portion of the papers were categorized the same each time, with some variation switching between the first and second categories. Table 2 displayes the confusion matrix for this classification. These results are extremely similar to organization based on citation, although using the uploaded papers increased true positive rate and decreased the false negative rate slightly. We found the same two papers which concern G-estimation continued to be incorrectly identified as relevant. However, there were two papers which were newly identified to be not relevant, although these papers are extremely clear in their relation to dWOLS, thus this appears to be a hallucination. 

              \begin{table}[H]
                \centering
                \begin{tabular}{|c|c|c|}
                \hline
                     & ChatGPT Relevant & ChatGPT Not Relevant \\
                     \hline
                 Relevant & 23 & 4\\
                    \hline
                 Not Relevant & 2 & 8\\
                    \hline
                \end{tabular}
                \caption{Confusion matrix for the classification of papers by ChatGPT using uploaded articles}
                \label{tab:placeholder}
            \end{table}

                \begin{itemize}
            \itemsep0em
                \item \textbf{True positive rate:} the probability that GenAI classified the paper as relevant and it actually is relevant, is 85\%
                \item \textbf{True negative rate:} the probability that GenAI classified the paper as not relevant and it actually is not relevant, is 80\%
                \item \textbf{False positive rate:} the probability that GenAI classified the paper as relevant when it actually is not relevant is 20\%
                \item \textbf{False negative rate:} the probability that GenAI classified the paper as not relevant when it actually is relevant is 15\%
            \end{itemize}

        \textbf{Create table:} We tasked ChatGPT to reproduce a table which appears in the Nadi review to categorize the papers to include our new found sources. The following are examples of prompts used: 

            \greybox{\textbf{Prompt:}
             I am writing a literature review on DTR estimation using dWOLS methodology introduced in “Doubly-robust dynamic treatment regimen estimation via weighted least squares” by Wallace and Moodie. I am looking for extensions dWOLS methodology and assessment/extension of dWOLS assumptions. Organize the uploaded pdfs into a table which includes columns: pdf file name, authors, title, contribution to literature review, and real-data analysis. 
             \\\textbf{Prompt:} 
            Create a table similar to the table on page 16 of (insert Nadi review 
            PDF file name) that uses all of the uploaded pdfs.}

        The resulting output contained numerous errors, including duplicate entries, incorrect classification, incorrect or missing real-data application, and other hallucinated details. ChatGPT also indicated that it could export the generated table to a PDF, but, this export did not function and no usable PDF file was produced.

\subsection{Summarize Sources} \label{summ}

This section covers the prompts and details from using GenAI to summarize given papers, using both the abstract and full-text. The included summaries were all sources from publicly available sources, such as abstracts and open-access articles.

\vspace{1em}
\textbf{Prompts:}
            \begin{itemize}
            \itemsep0em
                \item \textbf{Abstract-only summary:} I am writing a literature review on DTR estimation using dWOLS methodology introduced in Doubly-robust dynamic treatment regimen estimation via weighted least squares by Wallace and Moodie. I am looking for extensions dWOLS methodology and real data analysis. For the following paper, use only the abstract and provide a summary for the literature review.
            \\Paper: (insert paper citation)
            \\Abstract: (insert paper abstract)
            \item \textbf{PDF summary:} I am writing a literature review on DTR estimation using dWOLS methodology introduced in Doubly-robust dynamic treatment regimen estimation via weighted least squares by Wallace and Moodie. I am looking for extensions dWOLS methodology and real data analysis. Summarize the following paper and include subsections of Main Idea and Contribution, dWOLS Extension, Methodology, Applications, and a short paragraph summary for the literature review. Use only the pdf of the paper: (insert paper citation/pdf file name).
            \end{itemize}
            
        \subsubsection{Abstract summaries}

            \textbf{Structure:} Comparing the two GenAI models, one major difference in the results pertained to the structure of the summaries. Even though we did not specify a structure to adhere in our prompt, ScholarAI was entirely consistent throughout every summary response, including the sections: Context, Objective/Problem Addressed, Methodology, Real Data Application (if available), Conclusion. To see this structure, the following summaries were created by ScholarAI based on the abstract for ``Doubly Robust Estimation of Optimal Dosing Strategies" by Schulz \& Moodie (2020) (Summary 1) and ``Estimating Dynamic Treatment Regimes for Ordinal Outcomes with Household Interference" by Jiang et al. (2024) (Summary 2):
    
            \greybox{\textbf{Summary 1 (ScholarAI):}
            \\\textbf{Topic:} Doubly Robust Estimation of Optimal Dosing Strategies
            \\\textbf{Authors:} Schulz, J., \& Moodie, E. E. M. (2020)
            \\\textbf{Source:} Journal of the American Statistical Association, 116(533), 256–268. DOI:10.1080/01621459.2020.1753521
            \\\textbf{Context:} In precision medicine, tailoring treatment strategies to individual patient characteristics is paramount. While most existing methods, including the original dWOLS by Wallace and Moodie, focus on dichotomous treatments, there is a need to adapt these methods to continuous treatment settings, such as dosing strategies.
            \\\textbf{Objective:} This paper proposes an extension of the dynamic weighted ordinary least squares (dWOLS) framework to address continuous treatment scenarios, facilitating the estimation of optimal dosing strategies.
            \\\textbf{Methodology:}
            \\•	Extension of dWOLS: The proposed method extends the dWOLS framework to accommodate continuous treatments, offering a flexible regression-based estimation approach.
            \\•	Doubly Robust: The extension retains double robustness against model misspecification, provided the weights satisfy a specific balancing condition.
            \\•	Broad Applicability: A wide class of weight functions derived from this balancing condition allows for greater flexibility in modeling adaptive treatment strategies for continuous treatments.
            \\\textbf{Conclusion:} By extending the dWOLS methodology to continuous treatments, this study provides a robust and flexible framework for estimating optimal dosing strategies. This extension strengthens the application of dWOLS in precision medicine, allowing practitioners to tailor treatment more precisely to patient needs.
            \\This summary can be incorporated into your literature review to highlight the methodological advancements that extend dWOLS beyond dichotomous treatments, showcasing its application in developing optimal, individualized dosing strategies within precision medicine.}

            \greybox{\textbf{Summary 2 (ScholarAI):}
            \\\textbf{Topic:} Estimating Dynamic Treatment Regimes for Ordinal Outcomes with Household Interference
            \\\textbf{Authors:} Jiang, C., Thompson, M., \& Wallace, M. (2024)
            \\\textbf{Source:} Statistical Methods in Medical Research, 33(6), 981-995. DOI:10.1177/09622802241242313
            \\\textbf{Context:} Precision medicine often incorporates dynamic treatment regimes (DTRs) for decision support, which are pivotal in personalized healthcare strategies. While DTR estimation has been well-studied for numerous outcomes, its application to ordinal outcomes, particularly where interference is present, remains less explored.
            \\\textbf{Problem Addressed:} The study addresses the complexities of estimating DTRs for ordinal outcomes within contexts where treatment interference occurs among individuals in shared environments, such as households.
            \\\textbf{Methodology:}
            \\•	Weighted Proportional Odds Model: Introduces a regression-based approach that leverages an approximate doubly-robust method, accommodating ordinal outcomes.
            \\•	Interference Accounted: Utilizes covariate balancing weights based on joint propensity scores to model interference effects within household settings, demonstrating efficacy through simulation studies.
            \\•	Dynamic Weighted Proportional Odds Model: Extends the approach to estimate multi-stage DTRs with household interference.
            \\\textbf{Real Data Application:}
            \\•	Analysis: Applied the developed methodology to longitudinal data from the Population Assessment of Tobacco and Health study.
            \\•	Outcome Objective: Developed optimal treatment strategies to aid in household smoking cessation, considering interference within household units.
            \\\textbf{Conclusion:} This study advances the scope of dWOLS methodology by addressing ordinal outcomes and the challenge of treatment interference in shared environments, offering robust methodological enhancements for estimating individualized treatment regimes in complex real-world settings.
            \\This summary can be integrated into your literature review by emphasizing the extension of dWOLS methodology to accommodate ordinal outcomes and household interference, showcasing its applicability and robustness in diverse healthcare scenarios.}

            On the other hand, ChatGPT varied greatly in structure of response. For example, the summary of Schulz \& Moodie (2020) contains sections: What's new for dWOLS, Key idea, Flexibility, and Relevance to your review (Summary 3). To compare, the summary based on the abstract for ``Estimating Dynamic Treatment Regimes for Ordinal Outcomes with Household Interference" by Jiang et al. (2024) shows an entirely different approach, with sections: Extension relevant to dWOLS, Interference handling, Multi-stage generalization, Real-data application, and Takeaway for a dWOLS review (Summary 4).

            \greybox{\textbf{Summary 3 (ChatGPT):}\\
            Schulz \& Moodie (2020) — dWOLS extended to continuous treatments (optimal dosing)
            \\•   \textbf{What’s new for dWOLS:} Extends the Wallace–Moodie dWOLS framework from binary treatments to continuous doses, enabling estimation of optimal dosing strategies within a regression-based, dynamic treatment setting.
            \\•	\textbf{Key idea:} Introduces a balancing-condition–based weighting scheme; when the implemented weights satisfy this condition, the estimator is doubly robust to model misspecification.
            \\•	\textbf{Flexibility:} The balancing condition yields a broad class of valid weight functions, providing a flexible recipe for dosing-rule estimation across settings where treatment is continuously valued.
            \\•	\textbf{Relevance to your review:} Positions dWOLS as a foundation not only for binary decisions but also for dose optimization, preserving double robustness and expanding applicability to precision dosing problems. \\(No real-data application is described in the abstract.)}

            \greybox{\textbf{Summary 4 (ChatGPT):} \\
            Jiang, Thompson, \& Wallace (2024) — Ordinal-outcome DTRs with household interference
            \\•	\textbf{Extension relevant to dWOLS:} Proposes a regression-based, approximately doubly-robust framework—weighted proportional odds model—for single-stage DTR estimation with ordinal outcomes, conceptually aligned with dWOLS’s weighting-and-regression strategy.
            \\•	\textbf{Interference handling:} Accounts for household interference (one person’s treatment affecting another’s outcome) using covariate-balancing weights derived from joint propensity scores; verifies approximate double robustness via simulations.
            \\•	\textbf{Multi-stage generalization:} Extends to dynamic weighted proportional odds model for multi-stage DTRs under interference.
            \\•	\textbf{Real-data application:} Applies the methods to Population Assessment of Tobacco and Health (PATH) longitudinal survey data to deliver optimal household-level strategies for smoking cessation.
           \\• \textbf{Takeaway for a dWOLS review:} Demonstrates how dWOLS-style, weighted regression ideas can be adapted to ordinal endpoints and settings with interference, broadening practical DTR estimation beyond standard independent, continuous/binary outcomes.}

           Due to ScholarAI's structured formulaic responses which included contextual information, its summaries were more generally accessible for an audience who was not already familiar with the content. This consistency may also be beneficial for a researcher to compare and contrast key elements of multiple papers. 
           
           \textbf{Quality/Content:}  The depth of the summaries reflected the abstract. That is, if the abstract was missing information, the summary was also missing that information. In the summary based on the abstract for ``Doubly Robust Estimation of Optimal Dosing Strategies" by Schulz \& Moodie (2020), ChatGPT explicitly states that ``No real-data application is described in the abstract" (Summary 3). Although, the paper does include a real-data application, it was not discussed in the abstract, and thus, ChatGPT was unable to identify it.

           \textbf{Summarizing not relevant papers:} For papers that were not relevant to the literature review, we wondered if the GenAI models would recognize and state this lack of relevance, thus supporting the literature-screening process. However, we found that both models provided mixed results. For example, observe the abstract summary from both models for the paper ``Model Selection for G-Estimation of Dynamic Treatment Regimes" by Wallace, Moodie \& Stephens (2019). ChatGPT's summary states that this paper is centered on G-estimation, but still attempts to tie how this paper might be used in a literature review on dWOLS (Summary 5). ScholarAI explicitly confirms this paper regards G-estimation, but continues to discuss in the last paragraph how this summary can be integrated as a methodological advancement of dWOLS (Summary 6).

           \greybox{\textbf{Summary 5 (ChatGPT):}
           \\Wallace, Moodie, \& Stephens (2019) — Model selection for G-estimation of DTRs
            \\•	\textbf{Purpose:} Addresses a key barrier to applied DTR methods—choosing the blip (treatment-effect) model correctly—within the G-estimation framework.
            \\•	\textbf{Methodological contribution:} Using a quadratic approximation (inspired by iteratively reweighted least squares), the authors derive a quasi-likelihood for G-estimation and from it an information criterion for blip model selection. They outline the criterion’s theoretical properties.
            \\•	\textbf{Empirical evidence:} Demonstrated across simulation studies and with STAR*D data.
            \\•	\textbf{Relevance to a dWOLS-focused review:} Although centered on G-estimation, the paper delivers a principled model-selection strategy for the blip, directly informative for regression-based, doubly robust DTR approaches like dWOLS, where correct blip specification is also critical for valid, interpretable individualized rules.}

            \greybox{\textbf{Summary 6 (ScholarAI):}
            \\\textbf{Topic:} Model Selection for G-Estimation of Dynamic Treatment Regimes
            \\\textbf{Authors:} Wallace, M., Moodie, E., \& Stephens, D. (2019)
            \\\textbf{Source:} Biometrics, 75(4), 1205-1215. 
            \\\textbf{Context:} Dynamic treatment regimes (DTRs) are essential to personalized medicine, as they personalize treatment decisions based on individual patient characteristics. G-estimation is a method used to derive optimal DTRs by targeting the parameters of the blip function within a structural nested mean model. However, its implementation has been limited due to complexities in its specification and execution.
           \\\textbf{ Objective:} This paper aims to enhance G-estimation for DTRs by introducing a model selection framework that simplifies blip function specification, thereby facilitating its broader application.
            \\\textbf{Methodology:}
            \\•	Quadratic Approximation: Utilizes a quadratic approximation method, inspired by iteratively reweighted least squares, to create a quasi-likelihood function for G-estimation.
            \\•	Information Criterion: Develops an information criterion based on this quasi-likelihood to guide blip model selection, ensuring correct specification and improving accuracy in identifying optimal DTRs.
            \\\textbf{Evaluation:}
            \\•	Theoretical Properties: Discusses the theoretical foundations and properties of the proposed model selection criterion.
            \\•	Simulation Studies: Conducts simulation studies demonstrating the practical application and effectiveness of the criterion in various scenarios.
            \\•	Real Data Application: Applies the model selection approach to data from the Sequenced Treatment Alternatives to Relieve Depression (STAR*D) study, illustrating its real-world utility.
            \\\textbf{Conclusion:} This study proposes a novel model selection technique for G-estimation within DTR frameworks, leveraging quasi-likelihood and information criteria to improve model specification and application. This advancement potentially enhances the implementation of G-estimation in personalized medicine by overcoming previous practical barriers to its use.
            \\This summary can be integrated into your literature review to highlight methodological advancements in dWOLS and related frameworks, focusing on enhancing model selection processes critical for accurate and personalized treatment strategies.}

        \subsubsection{Full-text summaries} \label{summ_full}

            \textbf{Uploading papers:} ChatGPT and ScholarAI allow users to upload PDFs for the model to access and read, although both models displayed issues with this feature. ChatGPT has an upload limit related to the plan purchased. In some cases, it was unable to read the target PDF, or it read the incorrect PDF. However, ChatGPT clearly cites which paper it read in its response, so we adjusted prompts when necessary (e.g. providing the PDF file name in the prompt). ChatGPT encountered the most difficult when there were multiple papers with similar authors. ScholarAI does not have an upload limit, but it regularly struggled to upload and access some papers. There were a handful of papers which required to be re-uploaded in two or three new sessions for ScholarAI to be able to access it. Also, ScholarAI does not respond to general prompts to read the PDFs, instead, they require you to use their unique identifier ``PDF ID" which is assigned after uploading the paper. These additional steps in re-uploading papers and gathering the corresponding PDF IDs slowed down workflow tremendously.

            \textbf{Structure:}
            The prompt for the PDF summaries included a request for specific sections in the summary (Main Idea and Contribution, dWOLS Extension, Methodology, Applications, and a short paragraph summary for the literature review). Similar to previous observations, ScholarAI was entirely consistent throughout each response and adhered to the given sections. ChatGPT was also able to adhere to the given sections, which shows that some variability in responses can be controlled using more specific prompts. 

            \textbf{Quality/Content:} While both GenAI adhered to the given format, they still varied in terms of content. Every summary provided by ScholarAI was entirely text-based, providing a written explanation of the provided paper. However, some ChatGPT summaries opted to include equations and formulas that demonstrate the methodology of the paper. This is seen in the summary of the thesis paper ``Sensitivity analysis for unmeasured confounding in tailored treatment rules estimated by dWOLS" by Bahrampour, Schulz \& Moodie (2023) (Summary 7). This ChatGPT summary does not define the variables introduced or completely provides the necessary context. Consequently, it is unlikely to be useful for a reader who is not familiar with the base methodology. For instance, a general understanding regarding this topic is that $A$ represents treatment, but this is not clarified in the summary.

            \greybox{\textbf{Summary 7 (ChatGPT):}
            \\Bahrampour, Schulz \& Moodie (2023) — Sensitivity analysis for unmeasured confounding in tailored treatment rules estimated by dWOLS (Thesis)
            \\\textbf{Main Idea and Contribution}
            \\This thesis develops a practical sensitivity analysis for individualized/tailored treatment rules (ITRs) estimated via dWOLS when the no–unmeasured confounding (NUC) assumption may be violated. The central idea is to approximate the (unknown) impact of an unmeasured confounder $C$ using observed covariates $X$: write $E[C|A,X] \approx \eta_0+\eta_1 X$ and absorb its effect into an offset term in the outcome model. Varying a single sensitivity parameter (derived from $\eta_1$ and outcome coefficients) yields bias-adjusted blip estimates and, consequently, sensitivity-aware treatment rules. The work clarifies how violations of NUC distort dWOLS, proposes a simple corrective device, and evaluates it in simulations and a real-data application.
            \\ \textbf{dWOLS Extension}
            \\Standard one-stage dWOLS models the outcome as a treatment-free part plus a blip $\gamma(X,A;\psi)$ (the treatment contrast, typically linear in $A$ and $A:X$) and estimates $\psi$ by weighted least squares with balancing weights (e.g., absolute value weights or IPTW). The extension here keeps the dWOLS machinery intact but adds an offset $\alpha X$ to the regression, where
            $E[Y|A,X] \approx \alpha_0+\psi_0 A+\psi_1 (AX)+\alpha X,\alpha=\beta_x+\beta_c \eta_1.$
            \\Analysts probe a plausible range for $\alpha$ to assess robustness of the estimated blip and the resulting treatment rule to unmeasured confounding.
            \\ \textbf{Methodology}
                \\ • \textbf{Framework \& assumptions.} One-stage focus (two-stage explored in simulations), binary treatment, linear blip; dWOLS weights satisfy the balancing property so blip estimates are doubly robust when either the treatment model or treatment-free model is correct.
                \\ • \textbf{Derivation of the offset.} Start with an outcome model that depends on $X, A,$ and an unmeasured $C$. Replace $E[C|A,X]$ by a linear function of $X$ to obtain an approximating model with the offset $\alpha X$. This targets the dominant pathway by which $C$ confounds the $X–Y$ relation; it does not fully model the $C–A$ link, so some bias may remain.
                \\ • \textbf{Estimation algorithm (one stage).} Fit a treatment model to get propensities $\hat \pi (X)$ and compute balancing weights (absolute-value or IPTW). (1) Choose a grid for $\alpha$ informed by external data or subject-matter knowledge about $C–X$ dependence. (2) For each $\alpha $on the grid, fit weighted OLS of $Y$ on $A$ and $A:X$ with offset $\alpha X$ to obtain $\hat \psi (\alpha)$}

            Continuing, minor hallucinations were observed in the summaries, particularly with ScholarAI. In some cases, ScholarAI responded with summaries even when it was unable to read the PDF file. The following summary is an extremely vague recount of the abstract, which is supposed to be a summary of the paper ``Improved Doubly Robust Estimation in Learning Optimal Individualized Treatment Rules" by Pan \& Zhao (2020) (summary 8).

            \greybox{\textbf{Summary 8 (ScholarAI):}
            \\\textbf{Paper:} Pan, Y., \& Zhao, Y. Q. (2020). Improved Doubly Robust Estimation in Learning Optimal Individualized Treatment Rules. Journal of the American Statistical Association, 116(533), 283–294. https://doi.org/10.1080/01621459.2020.1725522
            \\\textbf{Main Idea and Contribution}
            \\The paper expands on the concept of doubly robust estimation in the context of learning optimal individualized treatment rules. The primary contribution is the enhancement of the doubly robust framework to improve estimation accuracy, thereby increasing the reliability and versatility of treatment rule frameworks. This advancement is critical for addressing complexities in clinical decision-making where individualized patient protocols are necessary.
            \\\textbf{dWOLS Extension}
            \\While specific extensions of dynamic Weighted Ordinary Least Squares (dWOLS) are not explicitly detailed in the initial chunk provided, the paper's focus on doubly robust methods suggests potential synergies with dWOLS methodologies, especially in optimizing treatment rule estimations through enhanced robustness tactics.
            \\\textbf{Methodology}
            \\The methodology centers around improving estimation techniques by employing advanced statistical methods to deepen the robustness of individualized treatment decision frameworks. The paper likely utilizes sophisticated modeling strategies that refine outcome predictions and adjust for potential confounding factors to deliver optimized treatment pathways based on individual patient data.
            \\\textbf{Applications}
            \\The applications of the methodologies discussed in the paper include scenarios where precise and personalized treatment regimens are imperative. Enhanced estimation processes are crucial in medical fields such as chronic disease management and personalized medicine, where patient-specific data drives clinical decisions.
            \\\textbf{Summary for Literature Review}
            \\By refining the doubly robust estimation approach, this paper makes significant strides in the realm of individualized treatment rule optimization. These advancements support the growing necessity for precision medicine techniques, ensuring that patient care strategies are both evidence-based and highly tailored. Such methodological improvements suggest future integration with existing frameworks like dWOLS, bolstering their applicability in diverse clinical settings.}

            Further prompts to attempt to get ScholarAI to produce a more accurate summary resulted in responses which indicated the model failed in retrieving the information needed for a summary (Response Excerpt 4).

            \greybox{\textbf{Response Excerpt 4:} Despite multiple attempts, extracting the needed detailed content for a comprehensive summary is proving challenging using the automated tools available. 
            }
            
            Continuing, both models occasionally left out crucial details of the paper in the summary. This was more commonly observed with ScholarAI. For example, in the summary for ``Doubly robust estimation of optimal dynamic treatment regimes with multicategory treatments and survival outcomes" by Zhang et al. (2022), an important part of the methodology is defining a new blip function, that is not tied to the reference treatment. This fact is entirely omitted by ScholarAI, and the methodology section only refers to comparing various weighting schemes. ChatGPT successfully mentions this details in its summary.

            Another example of this can be seen through the summaries for the paper ``Variable Selection in Regression-Based Estimation of Dynamic Treatment Regimes" by Bian et al. (2023). ChatGPT's summary discusses penalizing dWOLS using two penalty terms, and explicitly mentions the application to the STAR*D dataset (Summary 9). ScholarAI's summary does not mention these penalty terms, maintaining double robustness, or the exact dataset (Summary 10). Again, the lack of detail in this summary may be the result of difficulty in reading the paper.

            \greybox{\textbf{Summary 9 (ChatGPT):}
            \\Bian, Moodie, Shortreed \& Bhatnagar (2023) — Variable Selection in Regression-Based Estimation of Dynamic Treatment Regimes (Biometrics 79:988–999)
            \\\textbf{Main Idea and Contribution}
            \\The paper proposes penalized dWOLS (pdWOLS)—a variable-selection extension of the doubly-robust dWOLS framework—for learning optimal DTRs while automatically selecting (i) prognostic main effects and (ii) tailoring (treatment×covariate) interactions that define the decision rule. The authors motivate the need for principled selection because many time-varying features have weak prediction but strong decision value. pdWOLS keeps dWOLS’s simplicity (weighted regressions) and double robustness, adding penalties to perform estimation and selection simultaneously; simulations and an application to STAR*D illustrate performance. 
            \\\textbf{dWOLS Extension}
            \\pdWOLS augments dWOLS with two penalty terms—one for main effects and one for interactions—under a strong heredity constraint so that an interaction can be nonzero only if its corresponding main effects are nonzero. This is achieved by re-parameterizing interaction coefficients as $\psi_j = \psi_0 \tau_j \beta_j$ and minimizing a penalized weighted least-squares criterion; the approach preserves double robustness for the blip parameters (consistent if either the treatment model or the treatment-free model is correct). 
            \\\textbf{Methodology}
                 \\ • \textbf{Penalized objective.} Minimize $ Q(\theta)=L(Y;\theta)+\lambda(1-
                \alpha)||\beta||_1+\lambda \alpha ||\psi||_1$ (or elastic net/SCAD variants), with $\alpha$ controlling relative shrinkage of main vs interaction terms; the strong-heredity re-parameterization enforces structure in the fitted blip.
                \\ • \textbf{Computation.} A coordinate-descent/soft-thresholding scheme updates $\psi_0, \beta_j$, and the scaled interaction parameters via subgradient equations; updates use weighted residuals from partial fits, mirroring standard lasso machinery.
                \\ • \textbf{Asymptotics \& double robustness.} Under regularity, pdWOLS enjoys correct sparsity and asymptotic normality for active coefficients; when the blip is correctly specified, pdWOLS retains dWOLS’s double-robustness (consistency if either the treatment or treatment-free model is correct). 
                \\ •\textbf{Tuning \& practicalities.} Simulations use $\alpha$=0.5 and fourfold cross-validation for $\lambda$ ; main effects can be left unpenalized in low-dimensional settings to avoid dropping weak confounders. Competing baselines include Q-learning+LASSO and penalized A-learning (PAL).
            \\\textbf{Applications}
                \\ • \textbf{STAR*D trial (depression)}. Section 4 applies pdWOLS to STAR*D, demonstrating variable selection and rule estimation in a real clinical multi-stage setting. (The article flags the full analysis and results within this section.) 
                \\ • \textbf{Simulation studies.} The paper evaluates double-robustness, high-dimensional ((p $>$ n)) performance, and multistage settings, benchmarking pdWOLS against Q-learning+LASSO and PAL in terms of variable selection accuracy, decision error, and estimated value. 
           \\\textbf{Short paragraph for your literature review}
           \\Bian et al. (2023) extend dWOLS to pdWOLS, embedding lasso-style penalties (with strong heredity) into the weighted-regression framework so estimation and decision-focused variable selection occur together. The method preserves double robustness for the blip, has sparsity-selective asymptotics, and performs well in high-dimensional and multistage simulations against Q-learning+LASSO and PAL; a STAR*D application showcases real-data feasibility. For dWOLS-based DTR reviews, pdWOLS supplies a principled, interpretable way to identify who to tailor on while keeping the original method’s robustness and simplicity.}
            
            \greybox{\textbf{Summary 10 (ScholarAI):}
            \\ \textbf{Paper:} Bian, Z., Moodie, E., Shortreed, S., \& Bhatnagar, S. (2023). Variable Selection in Regression-Based Estimation of Dynamic Treatment Regimes. Biometrics, 79(2), 988-999. https://doi.org/10.1111/biom.13608
            \\ \textbf{Main Idea and Contribution}
            \\The paper focuses on a technique for improving dynamic treatment regimen (DTR) estimation through enhanced variable selection in regression-based methodologies. Its primary contribution is the introduction of systematic strategies for selecting variables that optimize the estimation of treatment rules, potentially reducing confounding effects and biases in complex datasets. The emphasis is on refining the regression models used to estimate DTRs, thereby enhancing personalization and effectiveness of the treatment strategies derived.
            \\ \textbf{dWOLS Extension}
            \\This paper extends the robustness of dWOLS methodologies by addressing a central issue of variable selection, which has implications for optimizing treatment decisions. The approach ensures that only pertinent variables are included in the decision-making framework, which aligns with the foundational principles of dwells in balancing bias and variance.
            \\ \textbf{Methodology}
            \\The authors employ regression-based selection mechanisms under a robust mathematical framework, integrating statistical techniques that fine-tune variable inclusions. This approach aims to maximize predictive accuracy and minimize errors associated with traditional regression models in treatment rule estimations. Various statistical models and simulations are utilized to validate the methodology and showcase its strength over conventional methods.
            \\ \textbf{Applications}
            \\The proposed framework is validated using simulated datasets that represent typical scenarios in personalized medicine where traditional methods may falter. By tailoring variable inclusion based on empirical evidence, the paper shows improvements in DTR estimates and consequent real-world applicability in fields like oncology, psychiatry, and chronic disease management.
            \\ \textbf{Literature Review Summary}
            \\ The paper provides significant advancements in the estimation of DTRs by optimizing variable selection processes within regression models, directly impacting methodologies like dWOLS. By ensuring only crucial variables inform treatment rules, the research enhances the precision of personalized medicine strategies, offering a substantial extension to existing DTR frameworks. This work will be invaluable in settings requiring nuanced treatment adaptations based on individualized patient data. }

            \textbf{Summarizing not relevant papers:}
            As in the abstract summaries, both models had success and failures when it came to recognizing that the paper should not be included in the literature review. For example, the paper ``Optimal dynamic treatment regime estimation in the presence of nonadherence" by Spicker et al. (2025) concerns G-estimation and non-adherence rather than dWOLS. The ScholarAI summary did not mention this fact, stating that it is an extension of dWOLS with nonadherence data. To compare, ChatGPT does not state that this is an extension of dWOLS, rather, that it ``broadens the settings in which dWOLS-style double-robust guarantees apply".

\subsection{Identify Gaps in Literature} \label{gap}

    The prompts used to request for gaps in literature were simple, but were repeated multiple times in sessions that had different contextual information. The scenarios were:  1. After requesting for abstract/PDF summaries; 2. After providing genAI with a list of citations; and 3. After providing genAI with uploaded full-text PDFs of the publications. The following are an example of some of the prompts which were used: 
    \begin{enumerate}
    \itemsep0em
        \item What are some gaps in literature for this topic?
        \item What are the most important gaps in literature for this topic?
        \item Refer to the uploaded documents and provide gaps in literature on the topic of DTR estimation using dWOLS.
    \end{enumerate}

    \subsubsection{ChatGPT}

            The gaps provided in the session with the abstract/PDF summaries were general and did not reference the works that discussed earlier in the session. A few gaps listed were topics that were, in fact, covered by the research summarized earlier in the session. For example, ChatGPT suggested extensions to ordinal outcomes and interference beyond dyads, both which are topics that appear in the provided literature.
            
            The context of providing citations and uploaded documents provided gap suggestions that were more focused and directly referenced the existing literature (Response Excerpt 5; note that the numbers in parenthesis are citations to the provided documents). While these gaps were more personalized than the previous attempts, there were still some identified which were covered in the given literature.

            \greybox{\textbf{Response Excerpt 5:}
            \\ 1) Outcome types and targets
            \\•	Beyond survival/ordinal/binary: You’ve got survival (19–22), binary (10), ordinal with interference (9). There’s little on counts, recurrent events, utility-based composites, or patient-reported outcomes formalized within dWOLS.
            \\•	Competing risks \& multi-state processes: Survival-focused dWOLS is present, but competing risks or illness–death models within dWOLS are not represented.}
            
            Another identified gap was “underrepresented domains” (Response Excerpt 5). Since the topic of DTR estimation is often applied to chronic conditions, suggesting applications in pediatrics shows that ChatGPT lacks insight that a researcher in this field might have.

            \greybox{\textbf{Response Excerpt 6:}
            \\ Underrepresented domains. Many apps in depression/RA/diabetes/HIV; fewer in oncology multi-line therapy, cardiology titration, perioperative care, and pediatrics, especially with continuous dosing or multi-drug choices.}
    
            We also prompted for the “most important gaps”, as it would be the job of a literature review to identify the most important and tangible gaps which could be the topic of next research. ChatGPT provided list of “high-priority” methodological gaps. One gap listed was “interference at scale” (Response Excerpt 7). A researcher who has understanding of the field would recognize that this gap is more of a data limitation rather than a methodological one, in which large-scale interference data is not common. Again, we observe that ChatGPT lacks the context a human researcher would have in identifying gaps in literature.
    
            \greybox{\textbf{Response Excerpt 7:}
            \\ Interference at scale (networks, households, dyads): Existing dWOLS extensions handle stylized exposures and small structures. We need scalable exposure learning (data-driven summaries), diagnostics for misspecified exposure mappings, and theory for partial/unknown networks—ideally with guarantees that persist under modest misspecification.}
            
        \subsubsection{ScholarAI}

            Like ChatGPT, ScholarAI was unable to provide suitable results when prompting in the same session as the abstract/pdf summaries. Also, the results from using a list of citations were broad and void of detail. The best performance was observed when the papers were uploaded to the model. 
            
            As before, ScholarAI was the most consistent between sessions, and the gaps provided were broader and sometimes too high-level. Some suggestions it gave was to integrate machine learning techniques, and to develop real-time clinical decision support tools based on dWOLS methodologies (Response Excerpt 8). While both of these are valid  potential research avenue, they are extremely high-level and not practical as starting points. 
            
            \greybox{\textbf{Response Excerpt 8:}
            \\ Integration of Emerging Machine Learning Techniques: While there are examples of integrating machine learning with dWOLS, such as using SuperLearner for improved confounding control, the literature still lacks comprehensive exploration of other advanced machine learning techniques, like deep learning, reinforcement learning, or unsupervised learning, and how they may enhance DTR estimates via dWOLS.
            \\ Real-time Decision Support Systems:
            There is a gap in developing real-time clinical decision support tools based on dWOLS methodologies, which would integrate seamlessly into health care systems for dynamic, live decision-making.}
            
            Another gap provided by the model was ``Generalizing Results Across Populations", which is vague and not dWOLS-specific. As well, the model suggested to combine dWOLS with casual inference techniques. dWOLS is already working with causal inference framework, so this suggestion is not appropriate given the setting. Like with ChatGPT, some of the gaps suggested by ScholarAI are lacking specific domain knowledge  (Response Excerpt 9).

            \greybox{\textbf{Response Excerpt 9:}
            \\ Generalizing Results Across Populations:
            Current methodologies may be tailored to specific population groups or data sets. Research aimed at generalizing results across diverse populations, potentially in multi-center or international studies, is essential.
            \\ Combination with Causal Inference Techniques: Although dWOLS is robust, combining it further with causal inference methodologies might enhance the causal interpretability and validity of the results, which is still an area in need of exploration.}

            Furthermore, when asked to provide the most important gaps in literature for the topic, ScholarAI listed "Integration with Machine Learning and AI" as first place, which again, might not be considered as the most actionable, important next-step in this research. 
            
\subsection{Create a Literature Review} \label{LR}

        Although ScholarAI advertises the ability to “create a literature review,” we were unable to obtain outputs resembling a coherent, report-style literature review, despite numerous attempts and prompting strategies. Thus, for this task, we focused on ChatGPT, specifically on the Thinking and Deep Research models.
        
        \subsubsection{GPT-5 Thinking}
            \greybox{\textbf{Prompt:} Write me a literature review on advancements in DTR estimation using dWOLS methodology introduced in “Doubly-robust dynamic treatment regimen estimation via weighted least squares” by Wallace and Moodie. I am looking for extensions dWOLS methodology and assessment/extension of dWOLS assumptions. Use the uploaded pdfs.}
            
            The initial literature review output from the normal Thinking model was not a complete, written report. The initial outputs always opted to use bullet points, and did not provide necessary background information on dWOLS that would be expected in a literature review. We repeatedly prompted ChatGPT to produce a report in paragraph form, but this output inherently lacked the flow and succinct organization of a real literature review, opting to include many sub-headings and lists. Despite this issue with style, the content covered a large portion of the provided papers, and no major hallucinations were observed. While papers and details were not explored to their entirety, this draft could serve as a reasonable starting point for a researcher.
            
            However, each literature review produced had a major issue with citations. In many instances, ChatGPT specifically talked about a paper and its contribution to dWOLS, but cited a completely different paper and author. For instance, ChatGPT cited "Doubly-Robust Dynamic Treatment Regimen Estimation with Binary Outcomes" by Cong Jiang, Michael Wallace, and Mary Thompson (2022) when it wrote about extensions to continuous treatments such as drug dosing, which should have been credited to "Doubly Robust Estimation of Optimal Dosing Strategies" by Juliana Schulz and Erica E. M. Moodie (2021). These errors are more subtle, and someone not completely familiar with all of the source material might overlook this. Interestingly, when requested to fix only the citations, ChatGPT still edited the content of the literature review, changing wording and organization of the written response.

        \subsubsection{GPT-5 Deep Research}

            \greybox{\textbf{Prompt:} Write me a literature review on advancements in DTR estimation using dWOLS methodology introduced in “Doubly-robust dynamic treatment regimen estimation via weighted least squares” by Wallace and Moodie. I am looking for extensions dWOLS methodology and assessment/extension of dWOLS assumptions}
            
            After being prompted, the Deep Research model asks clarifying questions to the user. For example: if the papers should be only peer-reviewed academic papers, what time frame should the literature review cover, if the review should focus on statistical methodology or applied settings, and if all the work should cite Wallace \& Moodie. 

            It is important to note that on its own, the Deep Research model is unable to access papers behind paywalls. Thus, in initial prompts to create the literature review, the model mainly re-stated the Nadi review plus information from additional open-access articles. We attempted to prompt the model to not consider the Nadi review, instead source from the uploaded papers. However, the model was unable to produce results solely on the uploaded sources, which is likely due to the model's nature to research in the internet. Regardless of this constraint, the Deep Research was able to write more effectively than the Thinking model and produce a report with improved flow and organization. Although, we still observed a lack of contextual information on dWOLS, which would be necessary for a literature review. While the Deep Research model did include information on the background of dWOLS, we observed an error: the model wrote that dWOLS assumes a Gaussian response, since it is a least squares model. This is incorrect, as we are not assuming our response variate is Gaussian when we are only estimating parameters. This small error is not obvious, but crucial in terms of background information. The Deep Research model also introduced terms and notation in the introduction which were not defined, which would be necessary context for the reader.
            
            Also, we noticed issues with citations, similar to the Thinking model where some authors were incorrectly attributed or not attributed at all. For example, citing "Dynamic Treatment Regimes on Dyadic Networks" by Marizeh Mussavi Rizi,  Joel A. Dubin, and Micheal P. Wallace (2024) for extension of dWOLS to ordinal outcomes, when it should be credited to "Estimating dynamic treatment regimes for ordinal outcomes with household interference: Application in household smoking cessation" by Cong Jiang, Michael Wallace, and Mary Thompson (2024). The model also forgot to include some necessary citations, such as on the topic of sensitivity analysis (Bahrampour et al., 2023). While the Deep Research model produced an impressive report, it still would require an expert in the subject matter to double-check for any mistakes, and fill in any gaps in contextual knowledge.

    \subsection{Use GenAI to write R Code based on papers} \label{code}

    \subsubsection{dWOLS}

    We attempted this task with different contexts, including the several combinations of uploading the dWOLS paper, and providing ChatGPT with the models (blip, treatment, treatment-free). The most basic model specification is identical to the one discussed in the dWOLS paper.

    \textbf{Prompts:}

    \begin{enumerate}
        \itemsep0em
        \item Write R code that implements the dWOLS function from the uploaded paper.
        \item Write R code that implements the dWOLS function from the uploaded paper for the following code that provides the data and models: \begin{verbatim}
            set.seed(1)
            # data setup
            n <- 10000
            # expit function
            expit <- function(x) {1/(1+exp(-x))}
            
            # model parameters a0, a1 (treatment model), p0, p1 (blip model)
            a0 <- 0; a1 <- 1; p0 <- 1; p1 <- 1
            
            # variables (X = patient information, A = treatment)
            X1 <- rnorm(n)
            A1 <- rbinom(n,1,expit(a0+a1*X1))
            
            X2 <- rnorm(n)
            A2 <- rbinom(n,1,expit(a0+a1*X2))
            
            # optimal treatments defined via blip functions
            A1opt <- as.numeric(p0 + p1*X1>0)
            A2opt <- as.numeric(p0 + p1*X2>0)
            
            # regrets - what is 'lost' by receiving non-optimal treatment
            mu1 <- (p0 + p1*X1)*(A1opt - A1)
            mu2 <- (p0 + p1*X2)*(A2opt - A2)
            
            # outcome assuming optimal treatment
            Yopt <- exp(X1) - X1^3 + exp(X2) - X2^3
            
            # outcome: optimal outcome minus regrets at each stage
            Y <- rnorm(n,Yopt - mu1 - mu2,2)
            
            # analysis
            # models to be passed to dWOLS
            y.mod <- Y~1
            blip.mod <- list(~X1,~X2)
            treat.mod <- list(A1~X1,A2~X2)
            tf.mod <- list(~X1,~X1+A1+X2)
            mydata <- data.frame(X1,X2,A1,A2,Y)
            k <- 2
        \end{verbatim}
    \end{enumerate}

    The code used to specify the component models is adapted from the supplementary material provided by Wallace \& Moodie (2015).

    \subsubsection{G-dWOLS}

    \textbf{Prompts:}
    \begin{enumerate}
        \itemsep0em
        \item Write code that uses the set-up from 3.1. Normally Distributed Doses in the uploaded paper. Use IPW weights, G-dWOLs, and output the psi parameters. Reference only the uploaded paper, and read the entire paper.
        \item Write code that uses the set-up from 3.5. Multiple Stages in the uploaded paper. Use IPW weights, G-dWOLS, and output the psi parameters for k=2 stages. Reference only the uploaded paper, and read the entire paper.
        \item Write R code to use G-dWOLS that uses the set-up from 3.5. Multiple Stages in the uploaded paper. Output the estimated psi parameters for k=2 stages. Reference only the uploaded paper.
        \item Implement G-dWOLS function from the uploaded paper, given this example data: \begin{verbatim}
            set.seed(1)
            # models to be passed to gdwols function
            # outcome
            outcome.mod<-y~1
            # blip model at each stage
            blip.mod<-list(~x1+a1,~x2+a2)
            # treatment model at each stage 
            treat.mod<-list(a1~x1,a2~x2)
            treat.mod.miss <-list(a1~x1,a2~1)
            # treatment-free model at each stage 
            tf.mod<-list(~log(x1)+sin(x1),~log(x2)+sin(x2))
            tf.mod.miss <-list(~x1,~log(x2)+sin(x2))
            
            # parameter settings
            # number of stages
            k<-2
            # number of bins for binned weight functions
            m<-5
            # sample size
            n<-1000
            
            # param values
            psi.mat<-matrix(rep(c(1,1,-1),2),byrow=TRUE,nrow=2)
            alpha.mat<-matrix(rep(c(-1,1),2),byrow=TRUE,nrow=2)
            
            # initialize output matrices
            psi.stage1<-NULL
            psi.stage2<-NULL
            
            # stage 1
            x1<-abs(rnorm(n,10,1))
            a1<-rnorm(n,alpha.mat[1,1]+alpha.mat[1,2]*x1,1)
            # stage 2
            x2<-abs(rnorm(n,10,1))
            a2<-rnorm(n,alpha.mat[2,1]+alpha.mat[2,2]*x2,1)
              
            # blips
            gamma1<-as.matrix(cbind(a1,a1*x1,a1^2)) %*% psi.mat[1,]
            gamma2<-as.matrix(cbind(a2,a2*x2,a2^2)) %*% psi.mat[2,]
            # y: outcome
            # y <- trmt free + blip
            y<-log(x1)+sin(x1)+log(x2)+sin(x2)+gamma1+gamma2 + rnorm(n,0,1)
            # data
            data<-data.frame(cbind(y,x1,x2,a1,a2))
        \end{verbatim}
    \end{enumerate}

    The code used to specify the component models is adapted from the supplementary material provided by Schulz \& Moodie (2020).



    \subsection{Writing R code and applying dWOLS to simulated data} \label{sim}

    \textbf{R code for simulation data:} adapted from supplementary material from Wallace \& Moodie (2015).
    \begin{verbatim}
        #--------------------------------------------------------------------------
        ## 1) simple simulation
        set.seed(1)
        
        # data setup
        n <- 1000
        # expit function
        expit <- function(x) {1/(1+exp(-x))}
        
        # model parameters a0, a1 (treatment model), p0, p1 (blip model)
        a0 <- -0.5; a1 <- 1;
        p0 <- -0.5; p1 <- 1;
        
        # variables (X = patient information, A = treatment)
        X1 <- rnorm(n) 
        X2 <- rnorm(n) 
        
        A1 <- rbinom(n,1,expit(a0+a1*X1)) # treatment stage 1
        A2 <- rbinom(n,1,expit(a0+a1*X2)) # treatment stage 2
        
        # optimal treatments defined via blip functions
        A1opt <- as.numeric(p0 + p1*X1>0)
        A2opt <- as.numeric(p0 + p1*X2>0)
        
        # regrets - what is 'lost' by receiving non-optimal treatment
        mu1 <- (p0 + p1*X1)*(A1opt - A1)
        mu2 <- (p0 + p1*X2)*(A2opt - A2)
        
        # outcome assuming optimal treatment
        Yopt <- 2*exp(X1) + X1^2 + 4*exp(X2) - X2^3
        
        # outcome: optimal outcome minus regrets at each stage
        Y <- rnorm(n,Yopt - mu1 - mu2,2)
        
        # create dataframe and export
        colnames(mydata) <- c("Risk1", "Risk2", "Treatment1", "Treatment2", "Health")
        write.csv(mydata, "health_data.csv", row.names=FALSE)
        
        # models to be passed to dWOLS
        y.mod <- Y~1
        blip.mod <- list(~X1,~X2)
        treat.mod <- list(A1~X1,A2~X2)
        tf.mod <- list(~X1,~X1+A1+X2)
        mydata <- data.frame(X1,X2,A1,A2,Y)
        k <- 2
        
        #--------------------------------------------------------------------------
        ## 2) complex simulation
        set.seed(1)
        
        # data setup
        n <- 1000
        # expit function
        expit <- function(x) {1/(1+exp(-x))}
        
        # model parameters a0, a1 (treatment model), p0, p1 (blip model)
        a0 <- -0.5; a1 <- 1; a2 <- 0.6;
        p0 <- -0.5; p1 <- 1; p2 <- 0.6;
        
        # variables (X = patient information, A = treatment)
        X1 <- rnorm(n) 
        A1 <- rbinom(n,1,expit(a0+a1*X1 + a2*X1^2)) # treatment stage 1
        
        X2 <- rnorm(n, mean = 0.3*X1, sd = 1) 
        A2 <- rbinom(n,1,expit(a0+a1*X2 + a2*X2^2)) # treatment stage 2
        
        # optimal treatments defined via blip functions
        A1opt <- as.numeric(p0 + p1*X1 + p2*X1^2>0)
        A2opt <- as.numeric(p0 + p1*X2 + p2*X2^2>0)
        
        # regrets - what is 'lost' by receiving non-optimal treatment
        mu1 <- (p0 + p1*X1 + p2*X1^2)*(A1opt - A1)
        mu2 <- (p0 + p1*X2 + p2*X2^2)*(A2opt - A2)
        
        # outcome assuming optimal treatment
        Yopt <- 2*exp(0.5*X1) + 1.5*X1^2 + 3*exp(0.3*X2) - X2^3 + 1.5*X1*X2
        
        # outcome: optimal outcome minus regrets at each stage
        Y <- rnorm(n,Yopt - mu1 - mu2,2)
        
        # create dataframe and export
        mydata <- data.frame(X1, X2, A1, A2, Y)
        colnames(mydata) <- c("Risk1", "Risk2", "Treatment1", "Treatment2", "Health")
        write.csv(mydata, "health_data2.csv", row.names=FALSE)
        
        # models to be passed to dWOLS
        y.mod <- Health ~ 1
        blip.mod  <- list(~ Risk1 + I(Risk1^2),
                          ~ Risk2 + I(Risk2^2))
        treat.mod <- list(Treatment1 ~ Risk1 + I(Risk1^2),
                          Treatment2 ~ Risk2 + I(Risk2^2) + Treatment1 + Risk1)
        tf.mod    <- list(~ Risk1 + I(Risk1^2),
                          ~ Risk1 + Treatment1 + Risk2 + I(Risk2^2))
        k <- 2
    \end{verbatim}
    
    \textbf{Prompts:}
    \begin{enumerate}
        \itemsep0em
        \item I have uploaded health\_data.csv which contains patient information for two stages of a DTR. In stage 1, we observed patient characteristic Risk1, prescribed Treatment1, and in stage 2 we observed patient characteristic Risk2, and prescribed Treatment 2. Treatment is a drug, where 1 means we give the patient the drug, and 0 means we give the patient a placebo. Finally, Health is the final measure of health for a patient, where higher values are better. Write R code to estimate the optimal DTR for this data using dWOLS, using the uploaded paper.
        \item I have uploaded health\_data.csv which contains patient information for two stages of a DTR. In stage 1, we observed patient characteristic Risk1, prescribed Treatment1, and in stage 2 we observed patient characteristic Risk2, and prescribed Treatment 2. Treatment is a drug, where 1 means we give the patient the drug, and 0 means we give the patient a placebo. Finally, Health is the final measure of health for a patient, where higher values are better. Use the uploaded paper biometrics.pdf to write R code estimate the optimal DTR using dWOLS for this data. Output the estimated coefficients. Use the following models:
        \begin{verbatim}
# models to be passed to dWOLS
y.mod <- Y~1
blip.mod <- list(~X1,~X2)
treat.mod <- list(A1~X1,A2~X2)
tf.mod <- list(~X1,~X1+A1+X2)
mydata <- data.frame(X1,X2,A1,A2,Y)
k <- 2
        \end{verbatim}
        \item	I have uploaded health\_data.csv which contains patient information for two stages of a DTR. In stage 1, we observed patient characteristic Risk1, prescribed Treatment1, and in stage 2 we observed patient characteristic Risk2, and prescribed Treatment 2. Treatment is a drug, where 1 means we give the patient the drug, and 0 means we give the patient a placebo. Finally, Health is the final measure of health for a patient, where higher values are better. Write R code to estimate the optimal DTR for this data using dWOLS and the DTRReg package.
        \item I have uploaded health\_data.csv which contains patient information for two stages of a DTR. In stage 1, we observed patient characteristic Risk1, prescribed Treatment1, and in stage 2 we observed patient characteristic Risk2, and prescribed Treatment 2. Treatment is a drug, where 1 means we give the patient the drug, and 0 means we give the patient a placebo. Finally, Health is the final measure of health for a patient, where higher values are better. How should I estimate the optimal DTR for this data?

    \end{enumerate}

    \subsection{Extend R package to G-dWOLS} \label{extendR}

    Prompts:
    \begin{enumerate}
        \itemsep0em
        \item Using the attached paper, extend the DTRreg function in the DTRreg R package to implement G-dWOLS. Include an example to run the function.
        \item Write code to extend the DTRreg function from the DTRreg package for multi-stage G-dWOLS
        \item Using the attached paper and files for the DTRreg R package, and edit the DTRreg function to implement G-dWOLS. Include an example to run the function.
    \end{enumerate}
    
\subsection{List of Papers Found by GenAI} \label{papers}
The following is the list of papers found by GenAI, and which GenAI found it.

\begin{enumerate}
\itemsep0em
\item Bahrampour, E., Schulz, J., \& Moodie, E. (2023). \textit{Sensitivity analysis for unmeasured confounding in tailored treatment rules estimated by dWOLS} (Master's thesis, McGill University). Retrieved from https://escholarship.mcgill.ca/concern/theses/wh246z332 (ScholarAI, ScholarAI GPT)

\item Bian, Z., Moodie, E. E. M., Shortreed, S. M., \& Bhatnagar, S. (11 2021). Variable Selection in Regression-Based Estimation of Dynamic Treatment Regimes. \textit{Biometrics, 79}(2), 988–999. doi:10.1111/biom.13608 (ChatGPT, ScholarAI, ScholarAI GPT)

\item Coulombe, J., Moodie, E. E. M., Shortreed, S. M., \& Renoux, C. (12 2020). Can the Risk of Severe Depression-Related Outcomes Be Reduced by Tailoring the Antidepressant Therapy to Patient Characteristics? \textit{American Journal of Epidemiology, 190}(7), 1210–1219. doi:10.1093/aje/kwaa260 (ChatGPT)

\item Coulombe, J., Moodie, E. E. M., Shortreed, S. M., \& Renoux, C. (2023). Estimating individualized treatment rules in longitudinal studies with covariate-driven observation times. \textit{Statistical Methods in Medical Research, 32}(5), 868–884. doi:10.1177/09622802231158733 (ChatGPT)

\item Danieli, C., \& Moodie, E. E. M. (2022). Preserving data privacy when using multi-site data to estimate individualized treatment rules. \textit{Statistics in Medicine, 41}(9), 1627–1643. doi:10.1002/sim.9318 (ChatGPT)

\item Dolmatov, M., Petrakos, N. Z., Moodie, E. E. M., Thomas, R., Durand, M., Klein, M. B., \& Pokomandy, A. de. (2024). Regression-Based Estimation of Optimal Adaptive Treatment Strategies: Key Methods. In D.-G. Chen (Ed.), \textit{Biostatistics in Biopharmaceutical Research and Development: Clinical Trial Analysis, Volume 2 }(pp. 363–389). doi:10.1007/978-3-031-65937-9\_12 (ScholarAI)

\item Dong, L., Moodie, E. E. M., Villain, L., \& Thiébaut, R. (2023). Evaluating the use of generalized dynamic weighted ordinary least squares for individualized HIV treatment strategies. \textit{The Annals of Applied Statistics, 17}(3), 2432–2451. doi:10.1214/22-AOAS1726 (ChatGPT, ScholarAI, ScholarAI GPT)

\item Hong, Y., Chen, L., Pan, Q., Ge, H., Xing, L., \& Zhang, Z. (2021). Individualized Mechanical power-based ventilation strategy for acute respiratory failure formalized by finite mixture modeling and dynamic treatment regimen. \textit{eClinicalMedicine, 36}. doi:10.1016/j.eclinm.2021.100898 (ChatGPT)

\item Jiang, C., Thompson, M., \& Wallace, M. (2024). Estimating dynamic treatment regimes for ordinal outcomes with household interference: Application in household smoking cessation. \textit{Statistical Methods in Medical Research, 33}(6), 981–995. doi:10.1177/09622802241242313 (ChatGPT)

\item Jiang, C., Wallace, M., \& Thompson, M. (2022). Doubly-Robust Dynamic Treatment Regimen Estimation for Binary Outcomes. \textit{arXiv [Stat.ME]}. Retrieved from http://arxiv.org/abs/2203.08269 (ChatGPT, ScholarAI, ScholarAI GPT)

\item Jiang, C., Wallace, M. P., \& Thompson, M. E. (2023). Dynamic treatment regimes with interference. \textit{Canadian Journal of Statistics, 51}(2), 469–502. doi:10.1002/cjs.11702
(ScholarAI, ScholarAI GPT)

\item Liang, D., Paul, A. K., Weir, D. L., Deneer, V. H. M., Greiner, R., Siebes, A., \& Gardarsdottir, H. (2025). Methods in dynamic treatment regimens using observational healthcare data: A systematic review. \textit{Computer Methods and Programs in Biomedicine, 263}, 108658. doi:10.1016/j.cmpb.2025.108658 (ChatGPT)

\item Liu, Y., Wang, Y., Kosorok, M. R., Zhao, Y., \& Zeng, D. (2018). Augmented outcome-weighted learning for estimating optimal dynamic treatment regimens. \textit{Statistics in Medicine, 37}(26), 3776–3788. doi:10.1002/sim.7844 (ScholarAI)

\item Ma, P., Liu, J., Shen, F., Liao, X., Xiu, M., Zhao, H., … Zhang, Z. (2021). Individualized resuscitation strategy for septic shock formalized by finite mixture modeling and dynamic treatment regimen. \textit{Critical Care, 25}(1), 243. doi:10.1186/s13054-021-03682-7 (ChatGPT)

\item Moodie, E. E. M., Coulombe, J., Danieli, C., Renoux, C., \& Shortreed, S. M. (2022). Privacy-preserving estimation of an optimal individualized treatment rule: a case study in maximizing time to severe depression-related outcomes. \textit{Lifetime Data Analysis, 28}(3), 512–542. doi:10.1007/s10985-022-09554-8 (ChatGPT)

\item Pan, Y., \& Zhao, Y.-Q. (2021). Improved Doubly Robust Estimation in Learning Optimal Individualized Treatment Rules. \textit{Journal of the American Statistical Association, 116}(533), 283–294. doi:10.1080/01621459.2020.1725522 (ScholarAI)

\item Rizi, M. M., Dubin, J. A., \& Wallace, M. P. (2024). Dynamic Treatment Regimes on Dyadic Networks. \textit{Statistics in Medicine, 43}(30), 5944–5967. doi:10.1002/sim.10278 (ChatGPT, ScholarAI, ScholarAI GPT)

\item Schulz, J., \& Moodie, E. E. M. (2021). Doubly Robust Estimation of Optimal Dosing Strategies. \textit{Journal of the American Statistical Association, 116}(533), 256–268. doi:10.1080/01621459.2020.1753521 (ChatGPT, ScholarAI, ScholarAI GPT)

\item Simoneau, G., Moodie, E. E. M., Nijjar, J. S., Platt, R. W., \& the Scottish Early Rheumatoid Arthritis Inception Cohort Investigators. (2020). Estimating Optimal Dynamic Treatment Regimes With Survival Outcomes. \textit{Journal of the American Statistical Association, 115}(531), 1531–1539. doi:10.1080/01621459.2019.1629939 (ChatGPT, ScholarAI GPT)

\item Simoneau, G., Moodie, E. E. M., Wallace, M. P., \& Platt, R. W. (2020). Optimal dynamic treatment regimes with survival endpoints: introducing DWSurv in the R package DTRreg. \textit{Journal of Statistical Computation and Simulation, 90}(16), 2991–3008. doi:10.1080/00949655.2020.1793341 (ChatGPT, ScholarAI, ScholarAI GPT)

\item Simoneau, G., Moodie, E., Azoulay, L., \& Platt, R. (2020). Adaptive Treatment Strategies With Survival Outcomes: An Application to the Treatment of Type 2 Diabetes Using a Large Observational Database. \textit{American Journal of Epidemiology, 189}(5), 461–469. doi:10.1093/aje/kwz272 (ChatGPT)

\item Simoneau, G., Moodie, E. E. M., Nijjar, J. S., \& Platt, R. W. (2020). Finite sample variance estimation for optimal dynamic treatment regimes of survival outcomes. \textit{Statistics in Medicine, 39}(29), 4466–4479. doi:10.1002/sim.8735 (ChatGPT)

\item Simoneau, G., Moodie, E., Platt, R., \& Chakraborty, B. (2017). Non-regular inference for dynamic weighted ordinary least squares: understanding the impact of solid food intake in infancy on childhood weight. \textit{Biostatistics, 19}(2), 233–246. doi:10.1093/biostatistics/kxx035 (ChatGPT, ScholarAI, ScholarAI GPT)

\item Spicker, D., \& Wallace, M. P. (2020). Measurement error and precision medicine: Error-prone tailoring covariates in dynamic treatment regimes. \textit{Statistics in Medicine, 39}(26), 3732–3755. doi:10.1002/sim.8690 (ChatGPT)

\item Spicker, D., Moodie, E. E. M., \& Shortreed, S. M. (2024). Differentially private outcome-weighted learning for optimal dynamic treatment regime estimation. \textit{Stat, 13}(1), e641. doi:10.1002/sta4.641 (ScholarAI GPT)

\item Spicker, D., Wallace, M. P., \& Yi, G. Y. (04 2025). Optimal dynamic treatment regime estimation in the presence of nonadherence. \textit{Biometrics, 81}(2), ujaf041. doi:10.1093/biomtc/ujaf041 (ChatGPT)

\item Talbot, D., Moodie, E. E. M., \& Diorio, C. (2023). Double robust estimation of optimal partially adaptive treatment strategies: An application to breast cancer treatment using hormonal therapy. \textit{Statistics in Medicine, 42}(2), 178–192. doi:10.1002/sim.9608 (ChatGPT, ScholarAI GPT)

\item Trejo Vargas, J. M. (2024). \textit{Novel strategies to address bias and variance in dynamic treatment regime models} (Doctoral dissertation). McGill University, Montréal, Québec, Canada. (ScholarAI, ScholarAI GPT)

\item Trenou, K. C., Mésidor, M., Eslami, A., Nabi, H., Diorio, C., \& Talbot, D. (2025). Using Machine Learning to Improve Control for Confounding in the Dynamic Weighted Ordinary Least Squares Estimator of Optimal Adaptive Treatment Strategies. \textit{Biometrical Journal, 67}(4), e70068. doi:10.1002/bimj.70068 (ChatGPT, ScholarAI, ScholarAI GPT)

\item Wallace, M. P., Moodie, E. E. M., \& Stephens, D. A. (2017). Model validation and selection for personalized medicine using dynamic-weighted ordinary least squares. \textit{Statistical Methods in Medical Research, 26}(4), 1641–1653. doi:10.1177/0962280217708665 (ChatGPT, ScholarAI, ScholarAI GPT)

\item Wallace, M., Moodie, E., \& Stephens, D. (2016). Model assessment in dynamic treatment regimen estimation via double robustness. \textit{Biometrics, 72}(3), 855–864. doi:10.1111/biom.12468 (ChatGPT, ScholarAI GPT)

\item Wallace, M., Moodie, E., \& Stephens, D. (2019). Model Selection for G-Estimation of Dynamic Treatment Regimes. \textit{Biometrics, 75}(4), 1205–1215. doi:10.1111/biom.13104 (ChatGPT, ScholarAI GPT)

\item Wang, C., \& Tom, B. D. M. (2025). A tutorial on optimal dynamic treatment regimes. \textit{arXiv [Stat.OT]}. Retrieved from http://arxiv.org/abs/2502.16988 (ScholarAI)

\item Yuxin, F., \& Moodie, E. E. M. (2016). \textit{Value search estimators of individualized treatment regimes using a new class of weights} (Master's thesis, McGill University). Retrieved from https://escholarship.mcgill.ca/concern/theses/rv042w606 (ScholarAI GPT)

\item Zhang, Z., Yi, D., \& Fan, Y. (2022). Doubly robust estimation of optimal dynamic treatment regimes with multicategory treatments and survival outcomes. \textit{Statistics in Medicine, 41}(24), 4903–4923. doi:10.1002/sim.9543 (ChatGPT, ScholarAI)

\item Zhao, Y.-Q., Zeng, D., Laber, E. B., \& Kosorok, M. R. (2015). New Statistical Learning Methods for Estimating Optimal Dynamic Treatment Regimes. \textit{Journal of the American Statistical Association, 110}(510), 583–598. doi:10.1080/01621459.2014.937488 (ScholarAI, ScholarAI GPT)

\end{enumerate}

Additional papers that were found (by ScholarAI), but were published before the original dWOLS paper (37, 38) or were entirely not relevant (39, 40):

\begin{enumerate}[resume]
\itemsep0em
    \item Chakraborty, B., \& Murphy, S. A. (2014). Dynamic Treatment Regimes. \textit{Annual Review of Statistics and Its Application, 1}(Volume 1, 2014), 447–464. doi:10.1146/annurev-statistics-022513-115553 (ScholarAI)

\item Zhao, Y. Q., \& Laber, E. B. (2014). Estimation of optimal dynamic treatment regimes. \textit{Clinical Trials, 11}(4), 400–407. doi:10.1177/1740774514532570 (ScholarAI)

\item Baldi Antognini, A., Frieri, R., Rosenberger, W. F., \& Zagoraiou, M. (2024). Optimal design for inference on the threshold of a biomarker. \textit{Statistical methods in medical research, 33}(2), 321–343. doi:10.1177/09622802231225964 (ScholarAI)

\item Zhou, Z., Teng, Z., Zhu, J., \& Tang, R. S. (2025). An improved biomarker-guided adaptive patient enrichment design for oncology trials. \textit{Journal of Biopharmaceutical Statistics, 35}(6), 1227–1243. doi:10.1080/10543406.2025.2489292 (ScholarAI)
\end{enumerate}

The one paper GenAI was unable to find was: 
\begin{enumerate}[resume]
    \item Zhang, Z., Zheng, B., \& Liu, N. (2020). Individualized fluid administration for critically ill patients with sepsis with an interpretable dynamic treatment regimen model. \textit{Scientific Reports, 10}(1), 17874. doi:10.1038/s41598-020-74906-z
\end{enumerate}

Categories:

\begin{table}[H]
\centering
\begin{tabular}{|c|c|l|p{4cm}|}
\hline
\multicolumn{1}{|l|}{}             & Category & \multicolumn{1}{c|}{Description} & \multicolumn{1}{c|}{Papers} \\ \hline
\multirow{3}{*}{Relevant}          & 1        &    In Nadi Review, Not found by GenAI                              &  41                           \\ \cline{2-4} 
                                   & 2        &    Found by GenAI, Not in Nadi Review                        &   1, 4, 5, 7, 9, 17, 22, 27,  28, 29, 34                          \\ \cline{2-4} 
                                   & 3        &    In Nadi Review and Found by GenAI                              &        2, 3, 8, 10, 11, 14, 15, 18, 19, 20, 21, 23, 24, 30, 35                     \\ \hline
\multicolumn{1}{|l|}{Not Relevant} & 4        &       Found by GenAI                           &              6, 12, 13, 16, 25, 26, 31, 32, 33, 36, 39, 40                \\ \hline
\end{tabular}
\caption{Papers found by GenAI sorted based on relevancy and if they were in the Nadi review. Papers 37 and 38 are excluded as they appeared in the Nadi review and were found by GenAI, but they concerned background information for methods that led to dWOLS, which we did not consider to be a relevant source. }
\label{table_cat}
\end{table}

\end{document}